\documentclass[9pt]{article}

\usepackage{graphics}
\usepackage{graphicx}
\usepackage{epsfig}
\usepackage{subfigure}
\usepackage{verbatim}
\usepackage{amsmath}
\usepackage{amssymb}
\usepackage{color}
\usepackage{bm}  
\usepackage{hyperref}
 \usepackage{stackrel}
 \usepackage{upgreek}
 \usepackage{csquotes}
 \usepackage{ulem}
 \usepackage{widetext}
 \usepackage{fullpage}

\DeclareGraphicsExtensions{.pdf,.eps,.png,.jpg,.mps}

\def\red{\textcolor{red}}

\newcommand{\SImum}{\ensuremath{\upmu}\textrm{m}\,}

\newcommand{\fraz}{\displaystyle\frac}
\def\tond#1{\left(#1\right)}

\def\quadr#1{\left[#1\right]}

\def\mod#1{\left|#1\right|}
\newcommand*\diff{\mathop{}\!\mathrm{d}}
\def\calE{{\cal E}}

\def\##1{{\bf #1}}
\def\=#1{\underline{\underline #1}}

\def\ux{\hat{\#x}}
\def\uy{\hat{\#y}}
\def\uz{\hat{\#z}}

\def\ko{k_{\scriptstyle o}}
\def\etao{\eta_{\scriptstyle o}}
\def\lambdao{\lambda_{\scriptstyle o}}
\def\lambdac{\lambda_{\rm c}}
\def\eps{\varepsilon}
\def\epso{\eps_{\scriptstyle o}}
\def\muo{\mu_{\scriptstyle o}}

\def\Nr{N_{\rm r}}
\def\Nmax{N_{\rm max}}

\def\muc{\mu_{\rm ch}}
\def\Edc{E_{\rm dc}}
\def\Bdc{B_{\rm 0}}

\def\Lsub{L_{\rm sub}}
\def\Lm{L_{\rm met}}

\def\epssub{\varepsilon_{\rm sub}}
\def\epsm{\varepsilon_{\rm met}}
\def\epsInSb{\=\eps_{\rm \InSb}}

\def\sigmagr{\sigma_{\rm gr}}
\def\taugr{\tau_{\rm gr}}

\def\SN{SiN$_{\rm x}$\,}
\def\InSb{InSb}
\def\NInSb{N_{\rm \InSb}}
\def\gammaInSb{\gamma_{\rm \InSb}}
\def\mstar{m_{\rm \InSb}}
\def\me{m_{\rm e}}
\def\qe{q_{\rm e}}

\def\kB{k_{\rm B}}
\def\qe{q_{\rm e}}
\def\epsdc{\eps_{\rm dc}}
\def\Vnm{V~nm$^{-1}$}
\def\THzK{~THz~K$^{-1}$}
\def\GHzK{~GHz~K$^{-1}$}
  
\def\THzT{~THz~T$^{-1}$}

\def\THzVnm{~THz~V$^{-1}$~nm}

\begin{document}

\begin{center}

\textbf{\Large Tricontrollable pixelated metasurface for absorbing terahertz radiation }\\
\vspace{4mm}

\textit{Pankaj Kumar}\footnote{Corresponding author: 
pankajjha@nitp.ac.in}\\
{Department of Electronics and Communication Engineering, National Institute of Technology Patna, Patna 800005, Bihar, India}
\vspace{4mm}

\textit{Akhlesh Lakhtakia}\\
{NanoMM---Nanoengineered Metamaterials Group,
Department of Engineering Science and Mechanics, The Pennsylvania State University, University Park, Pennsylvania 16802, USA}\\
{Material Architecture Center and Department of Electronics Engineering, Indian Institute of Technology (BHU), Varanasi 221005, Uttar Pradesh, India}\\
{Sektion for Konstruktion og Produktudvikling,  Institut for Mekanisk Teknologi, Danmarks Tekniske Universitet, 
DK-2800 Kongens Lyngby, Danmark}
\vspace{4mm}

 \textit{Pradip K. Jain}\\
 {Department of Electronics and Communication Engineering, National Institute of Technology Patna, Patna 800005, Bihar, India}\\
{Material Architecture Center and Department of Electronics Engineering, Indian Institute of Technology (BHU), Varanasi 221005, Uttar Pradesh, India}\\
\vspace{4mm}

\end{center}
 
\noindent\textbf{Abstract.}
The incorporation of materials with controllable electromagnetic constitutive parameters allows
the conceptualization and realization of controllable metasurfaces. With the aim of formulating and investigating a tricontrollable metasurface for efficiently absorbing terahertz radiation, we adopted a pixel-based approach in which the meta-atoms are biperiodic assemblies of discrete pixels. We patched some pixels with indium antimonide (InSb) and some
with graphene, leaving the others unpatched. The bottom of each meta-atom was taken to comprise a metal-backed substrate of silicon nitride. The InSb-patched pixels facilitate the thermal and magnetic control modalities, whereas the graphene-patched pixels facilitate the electrical control modality. With proper configuration of patched and unpatched pixels and with proper selection of
the patching material for each patched pixel, the absorptance spectrums of the pixelated metasurface were found to contain peak-shaped   features with maximum absorptance exceeding $0.95$, full-width-at-half-maximum bandwidth of  less than $0.7$~THz, and the maximum-absorptance frequency lying between $2$~THz and $4$~THz. The location of the  maximum-absorptance frequency can be thermally, magnetically, and electrically controllable. The  lack of rotational invariance of the optimal
meta-atom adds mechanical rotation as the fourth control modality. 

\vspace{4mm}

\section{Introduction}

A metasurface is conceptualized as a planar periodic array of identical, electrically thin meta-atoms whose lateral extent is a small fraction of the free-space wavelength $\lambdao$ \cite{Tong}.  {After relaxing the length restriction, the foregoing description applies even to many gradient metasurfaces \cite{Estakhri}
without any need to generalize the standard laws of reflection and refraction of plane waves \cite{Yu,Felbacq}.}
Although metasurfaces have been designed and fabricated for application in   spectral regimes ranging from   the  radio   to the visible  frequencies \cite{CTY,Glybovski}, they are particularly attractive for terahertz applications \cite{Siegel,Tao,Pawar} that include
imaging \cite{Yildrim}, spectroscopy \cite{Qin}, and cancer detection \cite{Yngvesson}. 

Absorption of terahertz radiation by metasurfaces is an area of ongoing research \cite{Thong,LiuFan,Huang,Liu-gold,Pankaj,Pankaj-err}. In many of these metasurface
absorbers, the meta-atom comprises graphene
on a dielectric substrate backed by a metal \cite{Thong,Huang,Pankaj}. The  frequency-dependent  surface conductivity
$\sigmagr$ of graphene  \cite{Depine} can be fixed by substitutional doping \cite{Castro}. More
importantly, $\sigmagr$
can be dynamically controlled by varying the chemical potential $\muc$ of graphene,    with the application of a quasistatic electric field \cite{Wang,Hanson-IEEE-AP2008} and/or a quasistatic magnetic  field \cite{Wright,Hanson-IEEE-AP2008}.  Accordingly, the spectrum of the  absorptance $A$ of a metasurface  containing graphene \cite{Huang,Pankaj} can be dynamically controlled. 

The application of a quasistatic electric field $\Edc$  (in \Vnm) along a fixed
direction provides a convenient control modality. For more reliable operation, an additional control modality would be desirable. At first glance, the
absolute  temperature $T$ (in K) appears promising as the second control modality because the temperature-dependent Fermi--Dirac distribution \cite{Kittel} is present in both
\begin{itemize}
\item the Kubo formula for $\sigmagr$ [Eq.~(1.34), Ref.~\cite{Depine}]
and
\item the Hanson formula for $\muc$ [Eq.~(53), Ref.~\cite{Hanson-IEEE-AP2008}].
\end{itemize}
However, the overall temperature-dependence is so weak for $T\in[273,323]$~K as to be inconsequential.  This means that a different material, with strongly temperature-dependent dielectric properties, must be
incorporated along with graphene in the meta-atoms.

Indium antimonide (InSb) is a promising material for this purpose. An isotropic $n$-type semiconductor,
InSb   has free carriers with low effective mass so that its conductivity can be thermally controlled \cite{Gruber,Zimpel}. When subjected to a quasistatic magnetic  field $\Bdc$ (in T) along a fixed direction,
InSb exhibits the uniaxial gyrotropic characteristics of a magnetoplasma \cite{Brion,Chen-book}. Thus,
InSb's magnetothermally dependent dielectric properties provide two additional control modalities in the terahertz regime \cite{HLTLW}.

{Parenthetically, multicontrollability abounds in nature. At different stages in our lives, we use different sets of muscles to move from one location to another. 
A message can be transmitted by one person to another by writing as well as by speaking; likewise, a message can be received by a person visually as well as aurally. In 
critical facilities---such as nuclear power plants, space stations, and missile guidance and command centers---multiple control modalities are used to reliably effect specific actions. As terahertz systems are being increasingly used for diverse applications, multicontrollability will be appreciated and even required in diverse scenarios.}

In this paper, we propose and simulate a tricontrollable metasurface   for efficiently
absorbing terahertz radiation. Each
meta-atom of this metasurface is made up of pixels \cite{Pankaj,Lakhtakia,Clemens,Chiadini}
affixed to a metal-backed dielectric substrate.  Some
pixels are patched with  \InSb, some with graphene, and the remaining ones
are unpatched. The pixellation of the metasurface
 permits trimodal control of the frequency of maximum absorptance.  The proposed absorber is also polarization insensitive.
 
The plan of this paper is as follows. Section~\ref{theory} provides descriptions of  the $T$- and $\Bdc$-dependences of the relative permittivity of InSb, the $\Edc$-dependence of the
surface conductivity of graphene, the
proposed metasurface,
and the boundary-value problem solved to determine
the absorptance of the metasurface when illuminated by a plane wave. Section~\ref{nrd}
provides a discussion of numerical results obtained by us. The paper concludes with
some remarks in Sec.~\ref{conc}.

An $\exp(-i\omega t)$ dependence on time $t$ is implicit, with $i=\sqrt{-1}$, $\omega=2\pi{f}$ as the angular frequency, and $f$ as the linear frequency. The free-space wavenumber
is denoted by $\ko=\omega\sqrt{\epso\muo}=2\pi/\lambdao$, where $\epso$ is the permittivity and $\muo$ is the
permeability of free space. Vectors are denoted by boldface letters;
the Cartesian unit vectors are denoted by 
$\ux$, $\uy$, and $\uz$; and dyadics are double underlined.


\section{Materials and Methods}\label{theory}

\subsection{Pixellated meta-atom}

\begin{figure}[ht]
\centering{\includegraphics[width=0.55 \columnwidth]{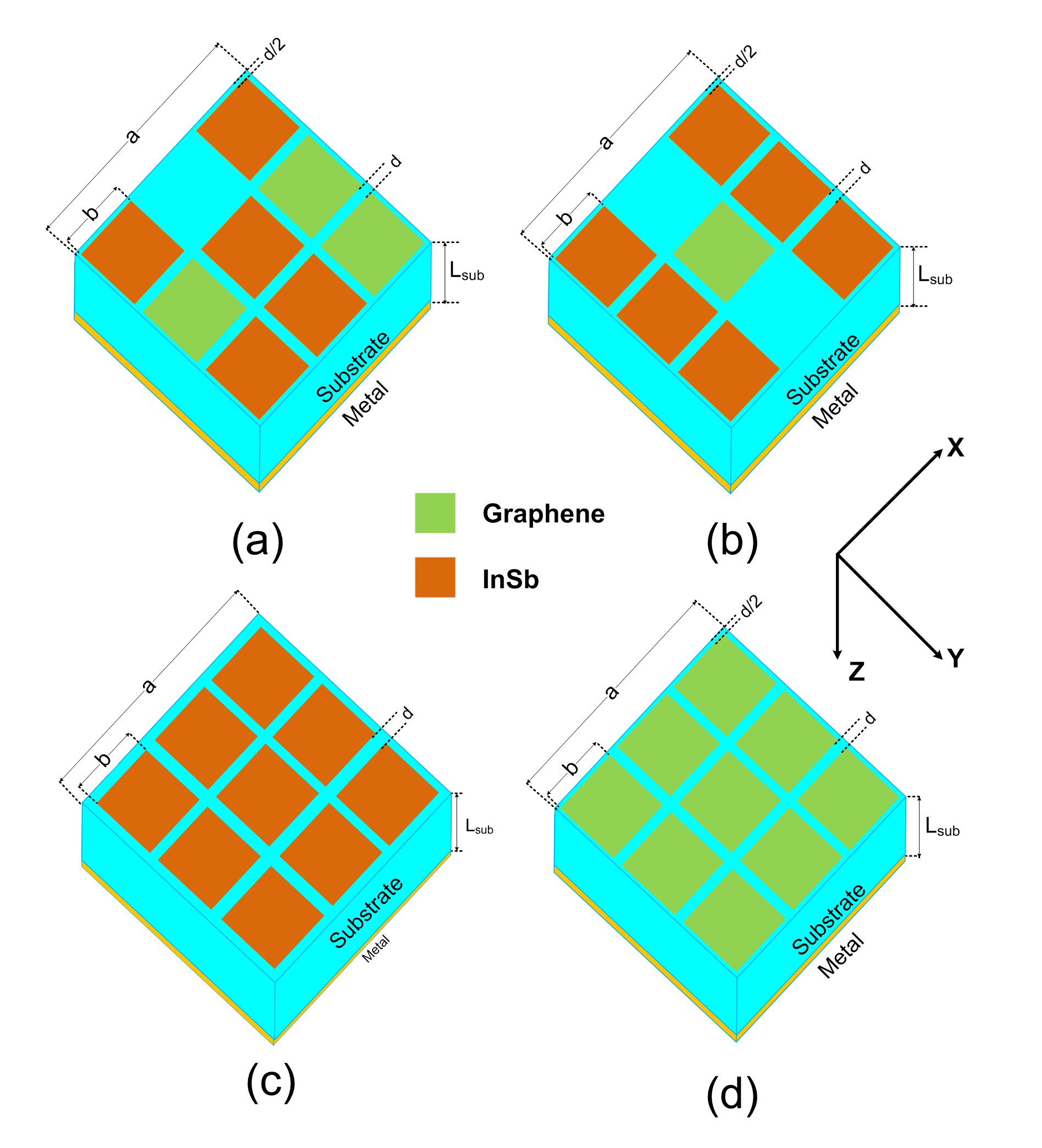}}
\caption{(a) Schematic of a meta-atom comprising $\Nmax=3\times3$ pixels, some patched with \InSb,
some patched with graphene, and the remaining unpatched,
on top of a metal-backed dielectric substrate. (b) Schematic of a meta-atom with one pixel patched with
graphene, six patched with \InSb, and two unpatched. (c and d) Schematics of meta-atoms with (c) all 
pixels patched with \InSb~and (d) all pixels patched with graphene.
\label{geometry}
}

\end{figure}

The metasurface is infinitely extended along the $x$ and $y$ axes, with the $z$ axis
normal to it.  The metasurface is biperiodic in the $xy$ plane, with each
each meta-atom   taken   to be a square of side   $a\leq\lambdac/4$ in
that plane, with $\lambdac$ being the lowest operational value of  $\lambdao$.
The bottom of each meta-atom
is a metal-backed dielectric   substrate, as shown in Fig.~\ref{geometry}.
The  thickness and   the relative permittivity scalar
of the substrate are denoted   by $\Lsub$  and $\epssub$, respectively; and those of the
metal by $\Lm$  and $\epsm$, respectively.

On top of the substrate in each meta-atom, an inner square of side $a-d$, $d\ll{a}$,
 is partitioned into square pixels of side $b$,
with each pixel separated from its nearest neighbor on every side by a
strip of thickness $d$. The dimensions $b$ and $d$ must be selected
so that the ratio $\Nr=a/(b+d)$ is an integer.
The  number of pixels in the meta-atom is thus $\Nmax=\Nr^2$.  The pixels are
aligned along the $x$ and $y$ axes. Some pixels
are patched with \InSb,
some patched with graphene, and the remaining ones are unpatched.

\subsection{Relative permittivity of InSb}

When a quasistatic magnetic field $\Bdc\ux$ is applied,
the relative permittivity dyadic $\epsInSb$ of \InSb~in the terahertz spectral regime can be stated
in matrix form as \cite{Chen-book}
\begin{equation}
\epsInSb=\begin{pmatrix} \eps_{\rm \InSb}^{\rm xx} & 0 & 0 \\ 0 & \eps_{\rm \InSb}^{\rm yy} & \eps_{\rm \InSb}^{\rm yz} \\ 0 & \eps_{\rm \InSb}^{\rm zy} & \eps_{\rm \InSb}^{\rm zz}\end{pmatrix}\,,
\end{equation}
where \cite{HLTLW}
\begin{equation}
\label{eq2}
\eps_{\rm \InSb}^{\rm xx}=
\eps_{\rm \InSb}^{(\infty)} - \qe^2\left(\frac{\NInSb}{\omega\epso\mstar}\right)
\frac{1}{\omega+i\gammaInSb} \,,
\end{equation}
\begin{eqnarray}
\nonumber
&&
\eps_{\rm \InSb}^{\rm yy}=
\eps_{\rm \InSb}^{\rm zz}= \eps_{\rm \InSb}^{(\infty)}
-\qe^2\left(\frac{\NInSb}{\omega\epso\mstar}\right)
\\[5pt]
&&\qquad 
\times
\frac{\omega+i\gammaInSb}
{\left({\omega+i\gammaInSb}\right)^2-\left(\frac{\qe\Bdc}{\mstar}\right)^2 }\,,
\label{eq3}
\end{eqnarray}
and
\begin{eqnarray}
\nonumber
&&
\eps_{\rm \InSb}^{\rm zy}=-\eps_{\rm \InSb}^{\rm yz}
=-i
\qe^2\left(\frac{\NInSb}{\omega\epso\mstar}\right) 
\\[5pt]
&&\qquad\times
\left(\frac{\qe\Bdc}{\mstar}\right)
\frac{1}
{\left({\omega+i\gammaInSb}\right)^2-\left(\frac{\qe\Bdc}{\mstar}\right)^2 }\,.
\label{eq4}
\end{eqnarray}
Here, $\eps_{\rm \InSb}^{(\infty)}=15.68$ is the high-frequency part, $\mstar=0.015 \me$,
$\me = 9.109 \times 10^{-31}$~kg is the   mass of an electron, $\qe=-1.602\times10^{-19}$~C
is the charge of an electron, $\gammaInSb = \pi\times 10^{11}$~rad~s$^{-1}$ is the damping
constant, 
\begin{equation}
\NInSb=5.76\times 10^{20}\,T^{3/2}\,\exp(-\calE_{\rm g}/2\kB T)
\label{eq5}
\end{equation}
is the temperature-dependent intrinsic carrier density (in m$^{-3}$), $\calE_{\rm g}=0.26$~eV
is the bandgap energy,
and
$\kB =8.617\times 10^{-5}$~eV~K$^{-1}$ is the Boltzmann constant \cite{Gruber,Zimpel,HLTLW}.

\begin{figure}[htb]
\centering{\includegraphics[width=0.55 \columnwidth]{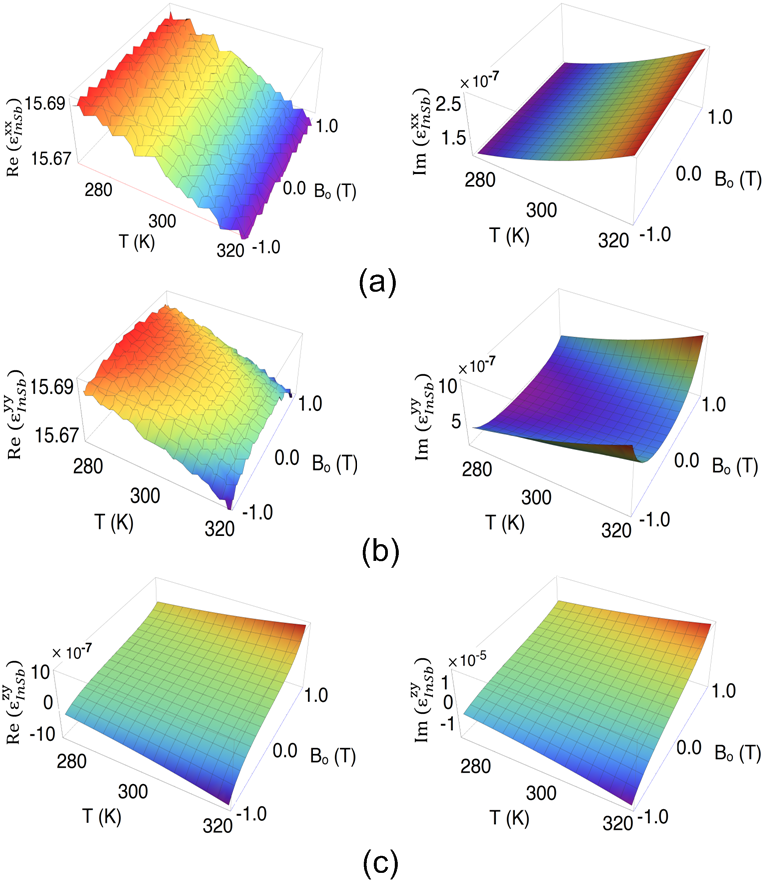}}
\caption{Real and  imaginary parts of (a) $\eps_{\rm \InSb}^{\rm xx}$, (b) $\eps_{\rm \InSb}^{\rm yy}$ ,
and (c) $\eps_{\rm \InSb}^{\rm zy}$
as functions
of $\Bdc\in[-1,1]$~T and $T\in[270,320]$~K. 
\label{InSb-relperm}
  }

\end{figure}

Components of $\epsInSb$ are shown as functions of $\Bdc\in[-1,1]$~T and $T\in[270,320]$~K
in Fig.~\ref{InSb-relperm}. As is clear from Eqs.~(\ref{eq2})--(\ref{eq4}),
 reversal of the sign of $\Bdc$ affects only the two off-diagonal components of $\epsInSb$.
 That effect is an interchange of $\eps_{\rm \InSb}^{\rm zy}$ and $\eps_{\rm \InSb}^{\rm yz}$, but it cannot have a major consequence   as the off-diagonal components are very small compared to the diagonal
components of $\epsInSb$.

\subsection{Surface conductivity of graphene}
As a  0.335-nm-thick layer, graphene is modeled  
as an infinitesimally  thin sheet with surface conductivity  [Eq.~(1.34),Ref.~\cite{Depine}]
\begin{eqnarray}
\nonumber
&&\sigmagr=-\fraz{q_e^2 \taugr\tond{1-i\omega\taugr}}{\pi\hbar^2}
\Bigg[\displaystyle\int_{-\infty}^{\infty}\fraz{\mod{\calE}}{\tond{1-i\omega\taugr}^2}\fraz{\partial F\tond{\calE,\muc}}{\partial\calE}\diff\calE
\\[10pt]
&&\qquad
+\displaystyle\int_{0}^{\infty}\fraz{F\tond{\calE,\muc}-F\tond{-\calE,\muc}}{\tond{1-i\omega\taugr}^2+4\taugr^2\calE^2/\hbar^2}\diff\calE\Bigg]\,,
\label{eq:sigma}
\end{eqnarray}
where the momentum relaxation time  $\taugr$ is assumed to be independent of the energy $\calE$, \begin{equation}
F\tond{\calE,\muc}=\quadr{1+\exp\tond{\fraz{\calE-\muc}{\kB T}}}^{-1}
\label{eq:FermiDirac}
\end{equation}
is the Fermi--Dirac distribution  function \cite{Kittel}, and $\hbar=6.582\times10^{-16}$~eV~rad$^{-1}$~s is the reduced Planck constant. 

Under the influence
of $\Edc\uz$, which is directed normally to graphene, the chemical potential $\muc$ is a solution
of the equation \cite[Eq.~(53),~Ref.]{Hanson-IEEE-AP2008}
\begin{equation}
\Edc=\fraz{\qe}{\pi\epso\epsdc\hbar^2\upsilon_{F}^2} \int_{0}^{\infty}{\calE\left[F(\calE,\muc)-F\left(\calE+2\muc,\muc\right)\right]}\diff\calE \,,
\label{muc-1}
\end{equation}
where $\upsilon_F$   is the Fermi speed of graphene and $\epsdc$ is the dc relative permittivity of the substrate. 
The integral on the right side of Eq.~(\ref{muc-1}) can be determined analytically to yield
\begin{equation}
\frac{\pi\epso\epsdc\hbar^2\upsilon_{F}^2\Edc}{\qe\kB^2T^2}=
{\rm Li}_2\left[-\exp\left(-\frac{\muc}{\kB T}\right)\right]-
{\rm Li}_2\left[-\exp\left(\frac{\muc}{\kB T}\right)\right]\,,
\label{muc-2}
\end{equation}
where ${\rm Li}_\nu(\zeta)$ is the polylogarithm function of order $\nu$ and argument $\zeta$ \cite{poly}.
The Newton--Raphson technique \cite{Jaluria} can be used to determine $\muc$ as a function
of $\Edc$. Reversal of the sign of $\muc$ reverses the sign of $\Edc$, according to Eqs.~(\ref{muc-1}) and (\ref{muc-2}).
Graphene is not affected significantly by $\Bdc\ux$, {because
that quasistatic magnetic field is wholly aligned in the plane containing the carbon atoms
\cite{Hanson-IEEE-AP2008}.}

\begin{figure}[htb]
\centering{
\includegraphics[width=0.35 \columnwidth]{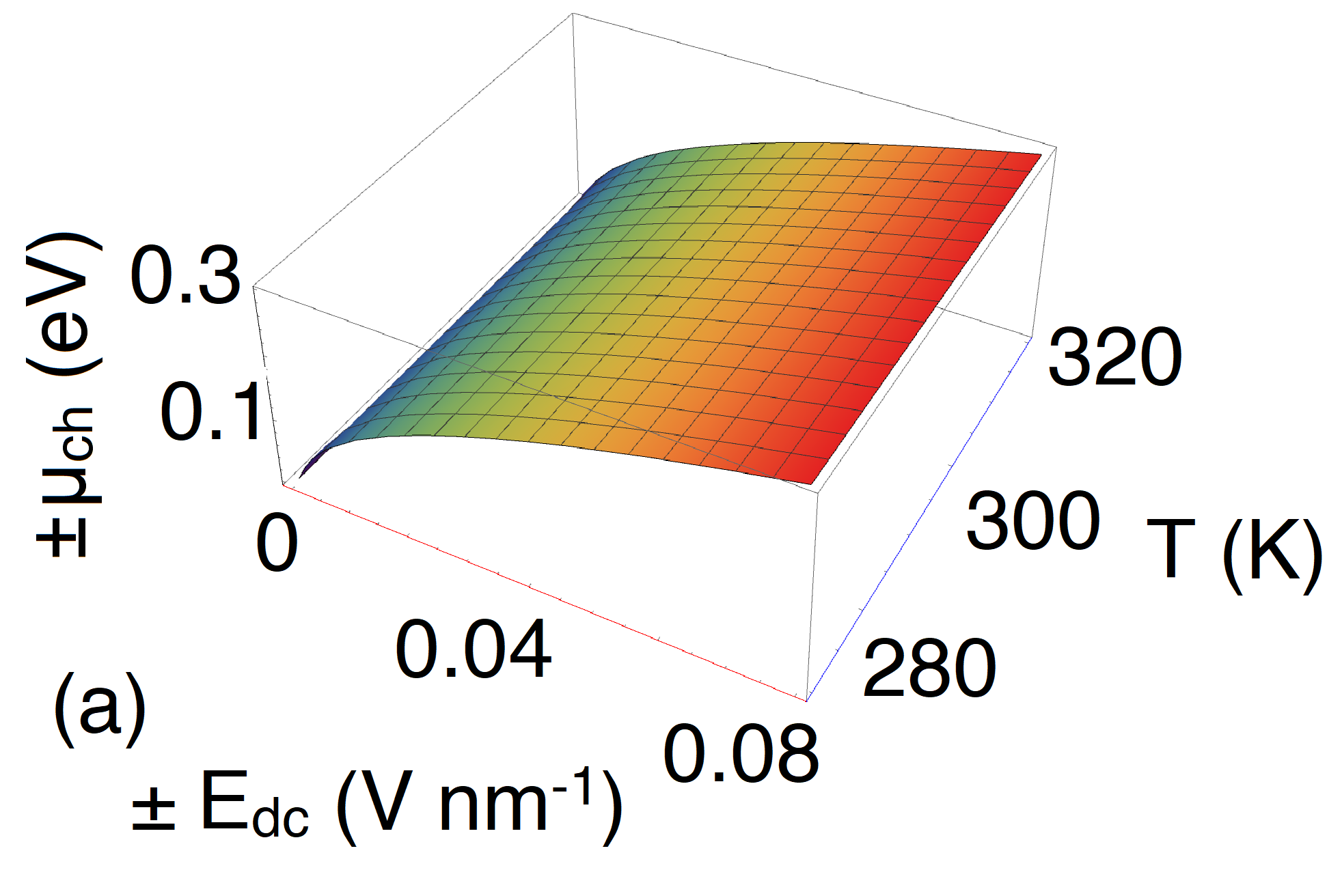}\\
\includegraphics[width=0.35 \columnwidth]{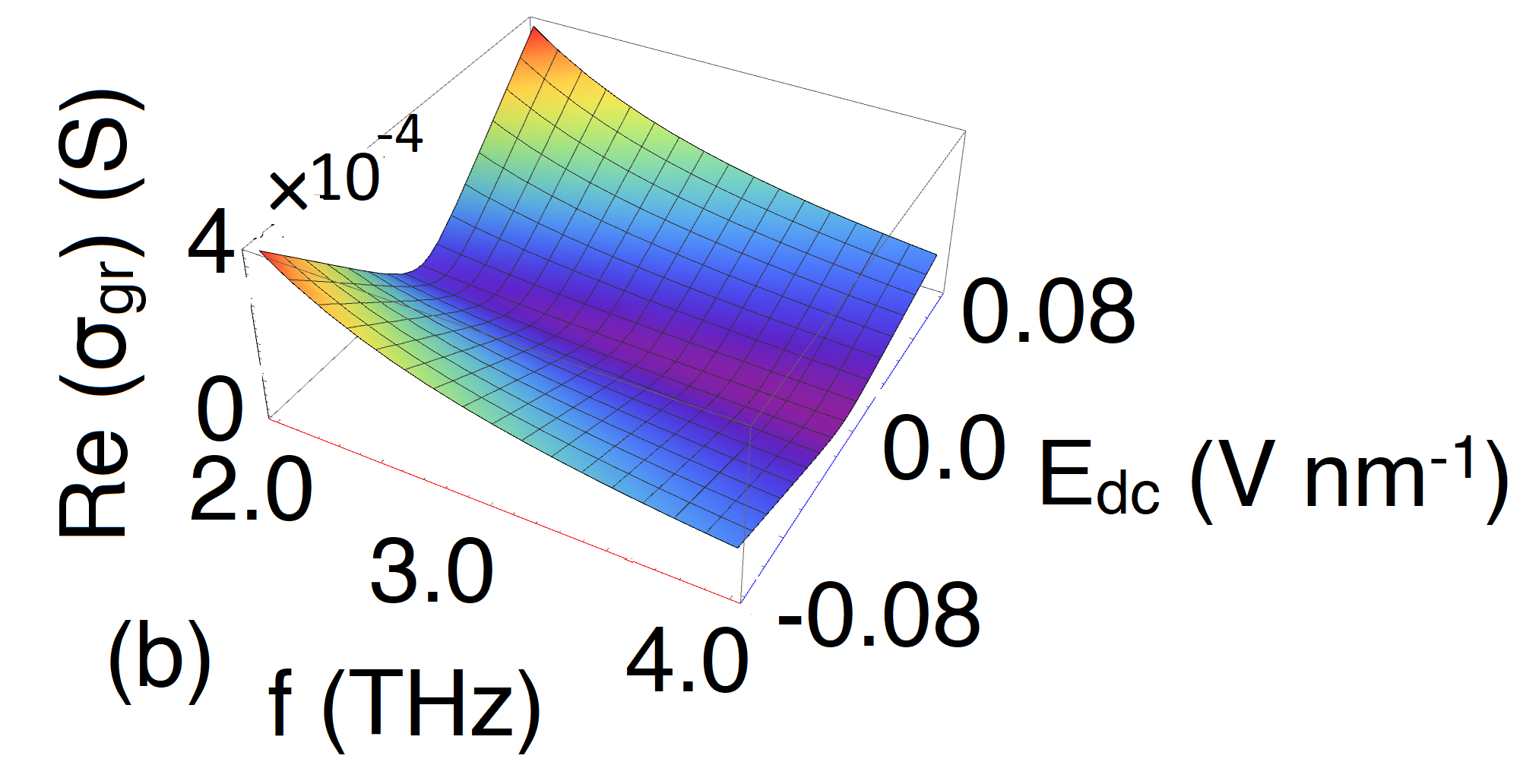}
\includegraphics[width=0.35 \columnwidth]{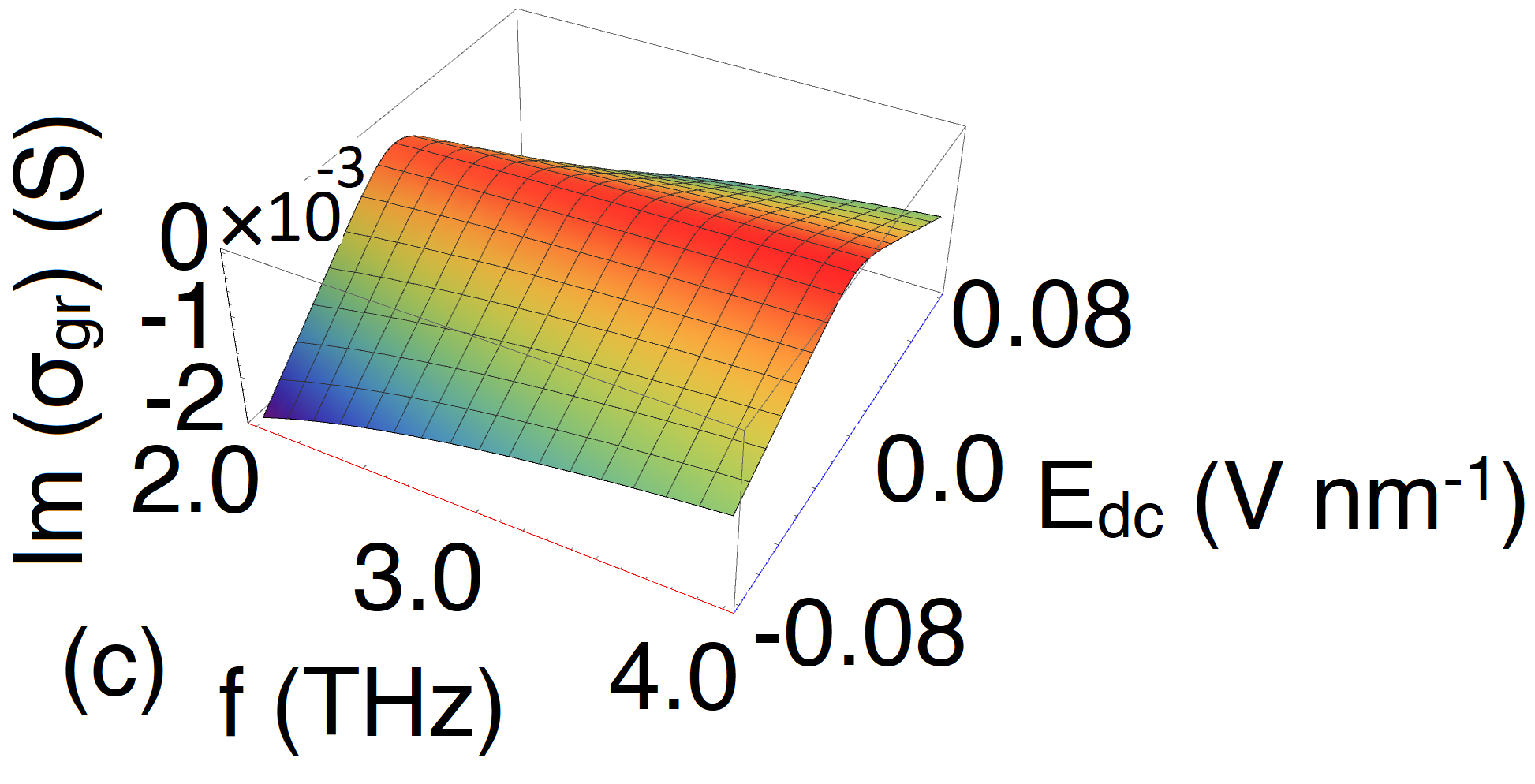}
}
\caption{(a) $\muc$ as a function of $\Edc\in[-0.08,0.08]$~\Vnm and $T\in[270,320]$~K,
when $\epsdc=7.5$. Note that the
signs of $\muc$ and $\Edc$ have to be the same.
(b) Real and (c) imaginary parts of $\sigmagr$  as functions of $f\in[2,4]$~THz and $E_{dc}\in[-0.08,0.08]$~\Vnm, when $\epsdc=7.5$ and $T=300$~K.
\label{Graphene-quants}
  }

\end{figure}

We fixed $\upsilon_F=10^{6}$~m~s$^{-1}$ \cite{Novoselov} and $\taugr= 1$~ps \cite{Wu}. Taking chemically prepared silicon nitride (SiN$_{\rm x}$) as the substrate \cite{Peralta}, we set $\epsdc=7.5$ \cite{SiNx-dc} and solved Eq.~(\ref{muc-2}) to determine $\muc$ as a function of $\Edc$ for $T\in\left\{270, 280, 290, 300, 310, 320\right\}$~K.  {Figure~\ref{Graphene-quants}a
shows $\muc$ as a   function of $\Edc\in[-0.08,0.08]$~\Vnm~and $T\in[270,320]$~K. Detailed examination of this figure reveals that
$\muc$ is practically independent of temperature for $T\in[270,320]$~K; thus, $\sigmagr$ is also temperature independent in the same temperature range.} As expected from Eq.~(\ref{muc-1}),
reversal of the sign of $\Edc$ reverses the sign of $\muc$ in Fig.~\ref{Graphene-quants}a.

Figures~\ref{Graphene-quants}b and \ref{Graphene-quants}c, respectively,
show the spectrums of the real and
imaginary parts of $\sigmagr$  in relation to $\Edc\in[-0.08,0.08]$~\Vnm~for $T=300$~K. 
Clearly, $\sigmagr$ is unaffected by the sign of $\Edc$.

\subsection{Plane-wave response}
The illumination of the pixelated metasurface   by a  normally incident, linearly polarized, plane wave was investigated.  The incident electric field phasor is then  given by
\begin{equation}
\#E_{\rm inc} = \left(   -\ux  \cos\varphi+\uy\sin\varphi \right)
\exp\left(i\ko z \right) \,
\label{E-inc}
\end{equation}
and the incident magnetic field phasor by
\begin{equation}
\#H_{\rm inc} =  -\etao^{-1}  \left(\ux  \sin\varphi+ \uy\cos\varphi \right)
\exp\left(i\ko z \right) \,,
\label{H-inc}
\end{equation}
where $\etao=\sqrt{\muo/\epso}$ is the intrinsic impedance of free space and $\varphi\in[0^\circ,90^\circ]$ is  the polarization angle. The plane wave is  transverse-electric \cite{Sadiku} or perpendicularly polarized \cite{BH1983} when $\varphi=90^\circ$ and transverse-magnetic \cite{Sadiku}  or parallel polarized 
\cite{BH1983} when $\varphi=0^\circ$. The substrate \SN was taken to be nondispersive and slightly dissipative in the 2--4-THz spectral regime; thus, $\epssub=7.6+i0.06$ \cite{SiNx-THz}. The metal backing the substrate was taken to be copper
\cite{Wu} with conductivity $5.9\times10^{7}$~S~m$^{-1}$, the thickness $\Lm=0.2$~\SImum being several times larger that the skin depth ($=0.046$~\SImum \cite{Sadiku}) in order
to prevent transmission through the metasurface.

In order to determine the reflected and transmitted electric and magnetic fields, a 3D
full-wave simulation in the 2--4-THz spectral regime
was carried out using the commercially available CST Microwave Studio\texttrademark~2019 software. Because of periodicity along the
$x$ and $y$ axes, the 2D Floquet model was used for the simulation. Adaptive    refinement of a mesh of tetrahedrons was done, with the number of tetrahedrons
being 47844 for the meta-atom depicted in Fig.~\ref{geometry}a,
16413 for Fig.~\ref{geometry}b,
17523 for Fig.~\ref{geometry}c,
and 52123 for Fig.~\ref{geometry}d. Being very thin, a graphene patch requires many more tetrahedrons than an \InSb~patch of the same lateral dimensions.

Transmission was found to be infinitesimal, as expected. The reflected electromagnetic field
contained both co-polarized and cross-polarized components, in general.
A post-processing module was used to compute the absorptance $A$, as explained
elsewhere \cite{LBG}, {by making use of the principle of conservation of energy and subtracting the sum of the co- and cross-polarized reflectances from unity.}

\section{Numerical Results and Discussion} \label{nrd}
Numerous different meta-atoms can be configured when $\Nr>2$ and the placement and the numbers
of InSb-patched, graphene-patched, and unpatched pixels are varied.  The temperature dependence of $\sigmagr$ is extremely weak for $T\in[270,320]$~K, graphene is unaffected
by the application of quasistatic magnetic field oriented tangentially to it,
whereas $\epsInSb$ is unaffected by $\vert\Edc\vert
\leq0.08$~\Vnm. 
However, $\sigmagr$ depends so strongly on $\Edc$ that even a few graphene-patched
pixels can overwhelm the magnetothermal control provided by the \InSb-patched pixels.

We fixed $a=9.6$~\SImum, $b=3$~\SImum,  $d=0.2$~\SImum, and $\Lsub=9.6$~\SImum, based on experience
\cite{Pankaj}. {After visually examining results obtained with all configurations possible
when $\Nr=3$, we settled on the configuration shown in Fig.~\ref{geometry}b:  
a single graphene-patched pixel accompanied by six InSb-patched  
and two unpatched pixels.} For comparison, we also calculated the spectrums of $A$ for meta-atoms with either  all nine InSb-patched pixels, as shown in Fig.~\ref{geometry}c, or all nine graphene-patched pixels, as shown in  Fig.~\ref{geometry}d.

\begin{figure}[htb]
\centering{
\includegraphics[width=0.35 \columnwidth]{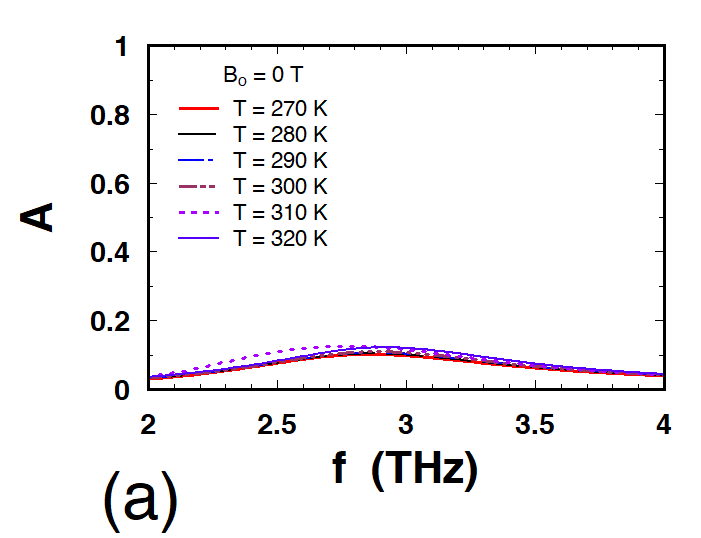}
\includegraphics[width=0.35 \columnwidth]{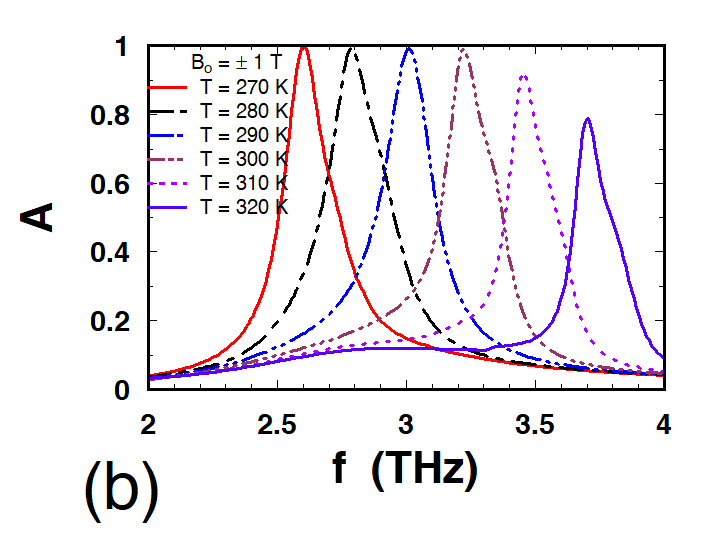}
\includegraphics[width=0.35 \columnwidth]{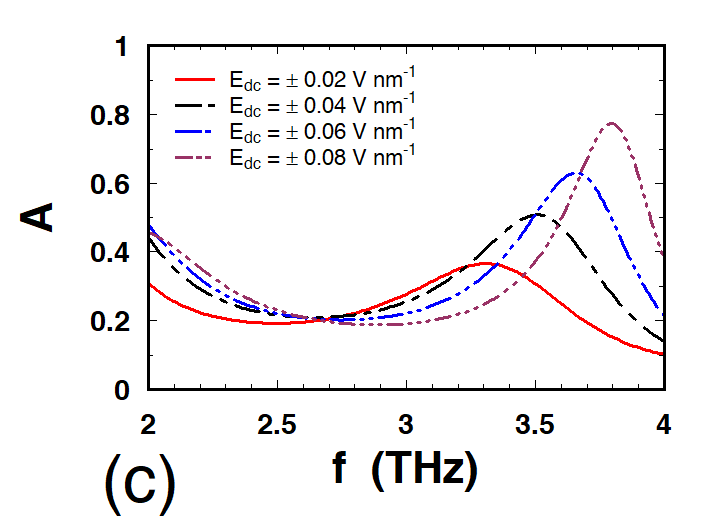}
}
\caption{(a and b) Absorptance spectrums of a metasurface with all nine pixels patched with
\InSb~(Fig.~\ref{geometry}c) for $T\in\left\{270, 280, 290, 300, 310,320\right\}$~K when {$\varphi=0^\circ$} and (a)
$\Bdc=0$ and (b) $\Bdc=\pm1$~T. (c) Absorptance spectrums of a metasurface with all nine pixels patched with
graphene~(Fig.~\ref{geometry}d) for $\Edc \in\left\{\pm0.02, \pm0.04, \pm0.06, \pm0.08\right\}$~\Vnm, when
$T=300$~K and  {$\varphi=0^\circ$}.
\label{Abs1}
  }

\end{figure}

\subsection{Thermal control modality}\label{tcm}
The thermal control modality is demonstrated in Fig.~\ref{Abs1}a containing the spectrums of
absorptance $A$ when all nine pixels are patched with \InSb~(Fig.~\ref{geometry}c), $\Bdc=0$, and 
$T$ changes from 270~K to 320~K in steps of 10~K, the value of $\vert\Edc\vert\leq0.08$~\Vnm~ being irrelevant.
The illuminating plane wave was taken to be normally incident with $\#E_{\rm inc}\parallel \ux$ for
 calculations.

Each of the six absorptance  spectrums  contains a very broad peak-shaped feature with a maximum  absorptance that lies between $0.10$ and $0.12$.  The maximum-absorptance frequency does change from
$2.86$~THz to $2.92$~THz at the rate of $1.2$~\GHzK~as the temperature increases from $270$~K to $320$~K, but the controllability provided by temperature alone is modest at best.  

The absorptance peaks narrow considerably and the maximum absorptances increase in 
Fig.~\ref{Abs1}b when $\vert\Bdc\vert$ is increased to $1$~T. Although the sign of $\Bdc$ does affect  $\eps_{\rm \InSb}^{\rm zy}$,  neither $\eps_{\rm \InSb}^{\rm xx}$ nor $\eps_{\rm \InSb}^{\rm yy}$ are affected by it. But, as the off-diagonal components of $\epsInSb$
are not of significant magnitude compared to the diagonal
components of $\epsInSb$, in the remainder
of this paper we have ignored the minuscule  dependence of $A$ on the sign of $\Bdc$.

As $T$ increases from $270$~K to $320$~K in Fig.~\ref{Abs1}b, the    maximum-absorptance frequency blueshifts at the rate of
 $\sim$0.02~\THzK from $2.59$~THz to $3.68$~THz and
 the maximum absorptance first increases slowly from $0.96$ to $0.985$ (at $T=290$~K)
 and then decreases rapidly to $0.69$ (at $T=320$~K). The full-width-at-half-maximum (FWHM) bandwidth lies between $0.19$ and $0.24$~THz for $T\leq300$~K, but drops to $0.15$~THz at $T=310$~K
 and $0.09$~THz at $T=320$~K. Thus, we conclude from Figs.~\ref{Abs1}a and \ref{Abs1}b that thermal control of \InSb-patched pixels should be more effective when a quasistatic magnetic field of sufficiently large magnitude is also applied.

 \subsection{Magnetic control modality}\label{mcm}
The magnetic control modality {due to InSb-patched pixels} is revealed by a comparison of Figs.~\ref{Abs1}a and \ref{Abs1}b. Increase of $\vert\Bdc\vert$ from $0$ to $1$~T
 causes the maximum-absorptance frequency to
 \begin{itemize}
 \item[(i)]  redshift by  $0.27$~THz at $T = 270$~K and
$0.09$~THz at $T = 280$~K, and
\item[(ii)] blueshift by
$0.10$~THz at $T = 290$~K, 
$0.31$~THz at $T = 300$~K,   
$0.53$~THz at $T = 310$~K, and 
$0.76$~THz at $T = 320$~K.
\end{itemize}
Simultaneously, the increase of $\vert\Bdc\vert$ enhances the maximum absorptance from low
 values ($0.11\pm0.01$) to values exceeding $0.68$ and as high as $0.985$. A reversal of the sign of $\Bdc$ is infructuous, for the reasons discussed in Sec.~3.\ref{tcm}.

\subsection{Electrical control modality}\label{ecm}
The electrical control modality is demonstrated in Fig.~\ref{Abs1}c containing the absorptance spectrums when all nine pixels are patched with graphene
(Fig.~\ref{geometry}d) and $\vert\Edc\vert$
changes from $0.02$~\Vnm~to $0.08$~\Vnm~in steps of $0.02$~\Vnm, the values
of $\Bdc$ and $T\in[270,320]$~K being irrelevant. The calculations were made
for a   normally incident plane wave with $\#E_{\rm inc}\parallel \ux$.

As is clear from Figs.~\ref{Graphene-quants}b and ~\ref{Graphene-quants}c,
reversal of the sign of $\Edc$  does not affect $\sigmagr$; hence, $A$ is unaffected
by the sign of $\Edc$.
Each of the four absorptance spectrums in Fig.~\ref{Abs1}c
contains a prominent peak-shaped feature. As $\vert\Edc\vert$ increases from
$0.02$~\Vnm~to $0.08$~\Vnm, the FWHM bandwidth decreases, the   maximum-absorptance frequency
 blueshifts from $3.29$~THz to $3.79$~THz
 at the rate of about
 $8.3$~THz~V$^{-1}$~nm, and
the maximum absorptance
rises from $0.35$ to $0.78$.
 The modest values of the maximum absorptance are
not surprising because graphene by itself is a modest absorber of low-THz radiation \cite{Kaipa}, although properly designed graphene metasurfaces can deliver almost the maximum absorptance possible \red{ \cite{Pankaj, PC, Rah}}.

\subsection{Tricontrollable metasurface}
Having confirmed the bicontrollability of InSb-patched pixels in Secs.~3.\ref{tcm}
and 3.\ref{mcm}
and the unicontrollability of graphene-patched pixels in Secs.~3.\ref{ecm}, we now move on to the metasurface depicted in Fig.~\ref{geometry}b: each   meta-atom has its central pixel patched with graphene, six pixels patched with \InSb, and two unpatched pixels.

The absorptance $A$ was calculated
for a   normally incident plane wave with $\#E_{\rm inc}\parallel \ux$. Figures~\ref{Abs2}a--i show 
$A$ as a function of  $f\in[2,4]$~THz and $T\in[270,320]$~K for nine combinations
of $\vert\Bdc\vert$ and $\vert\Edc\vert$.

\subsubsection{$\Bdc=0$}

Let us begin with $\Bdc = 0$. Figures~\ref{Abs2}a--c show a peak-shaped feature
in the absorptance spectrum,
with the maximum absorptance lying between  $0.$93 and $0.99$ and the
FWHM bandwidth between  $0.28$~THz and $0.48$~THz. Clearly,
thermal control provided by the six InSb-patched pixels is weak, the    maximum-absorptance frequency blueshifting at the rate of
about $0.0012$~\THzK. The weakness of thermal control in the absence of 
 a quasistatic magnetic field follows from Sec.~3.\ref{tcm}. 
 
 However,
 the electrical control modality provided by the sole graphene-patched
 pixel is strong in Figs.~\ref{Abs2}a--c. As $\vert\Edc\vert$ increases from
$0.02$~\Vnm~to $0.08$~\Vnm, the FWHM bandwidth decreases
and the   maximum-absorptance frequency
 blueshifts 
 at the rate of  
about $18.66$~ \THzVnm.
 
 \subsubsection{$\vert\Bdc\vert>0$}
 
\begin{widetext}
 \begin{figure}[ht]
\centering{
\includegraphics[width=0.3 \columnwidth]{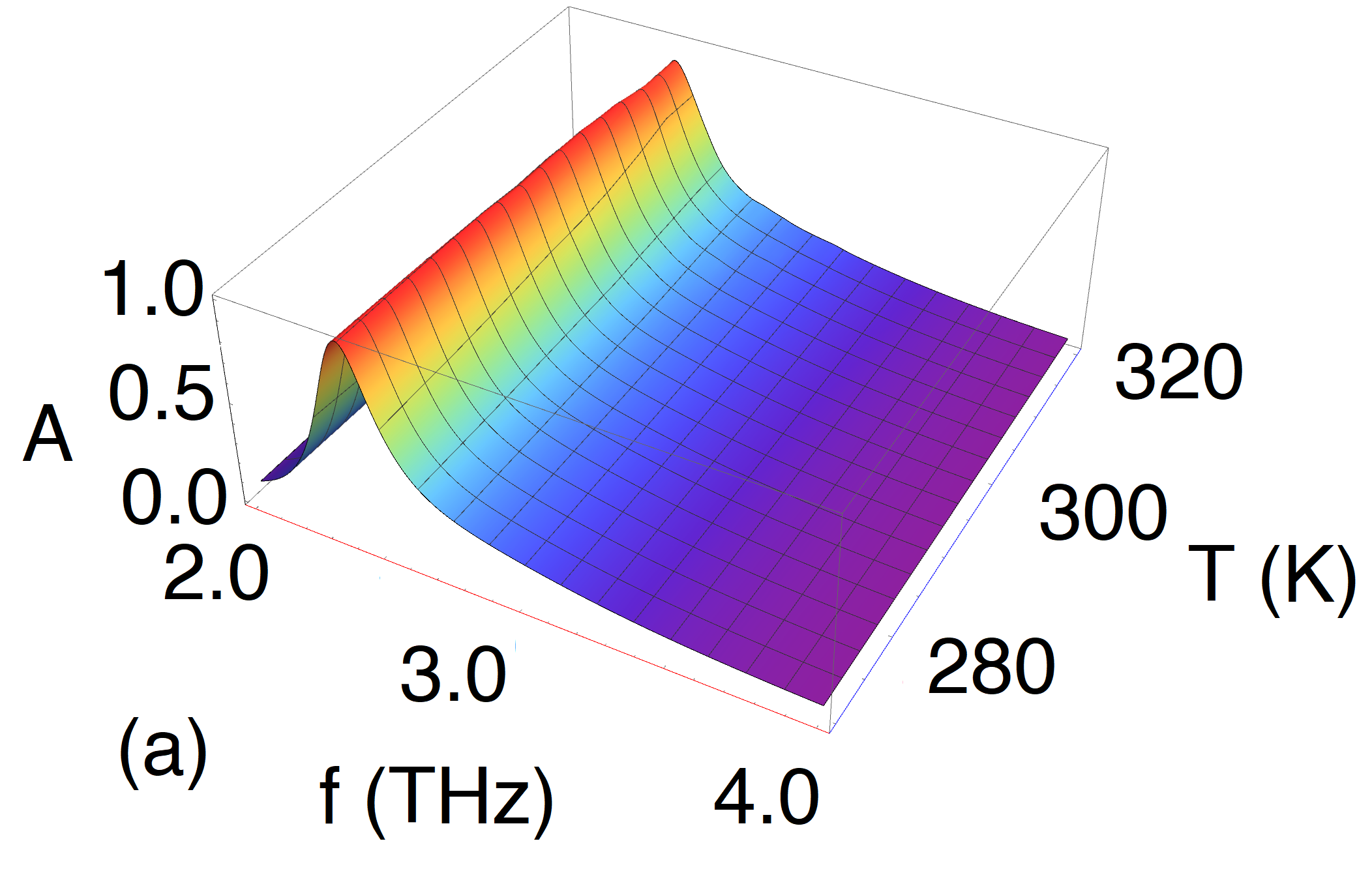}
\includegraphics[width=0.3 \columnwidth]{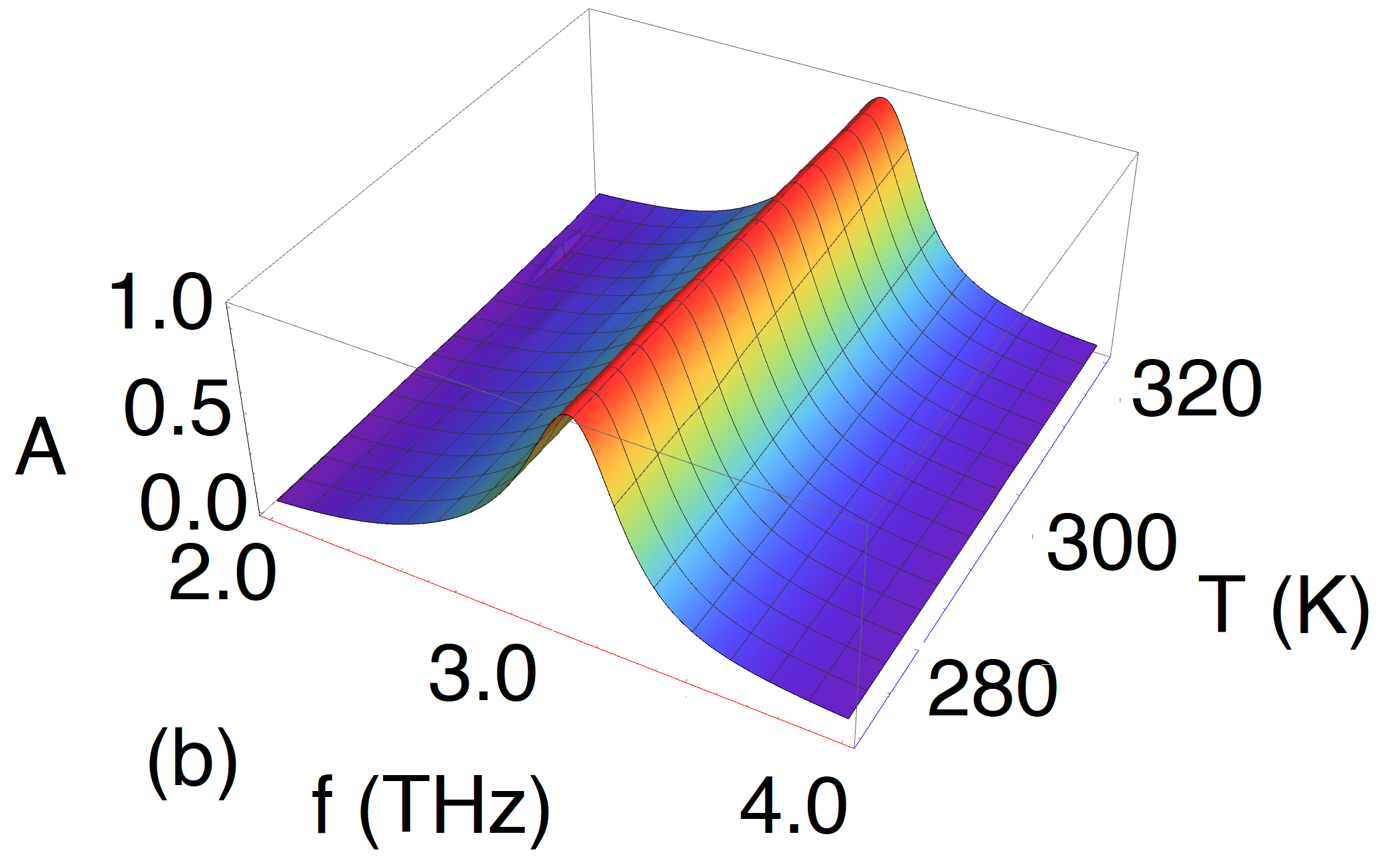}
\includegraphics[width=0.3 \columnwidth]{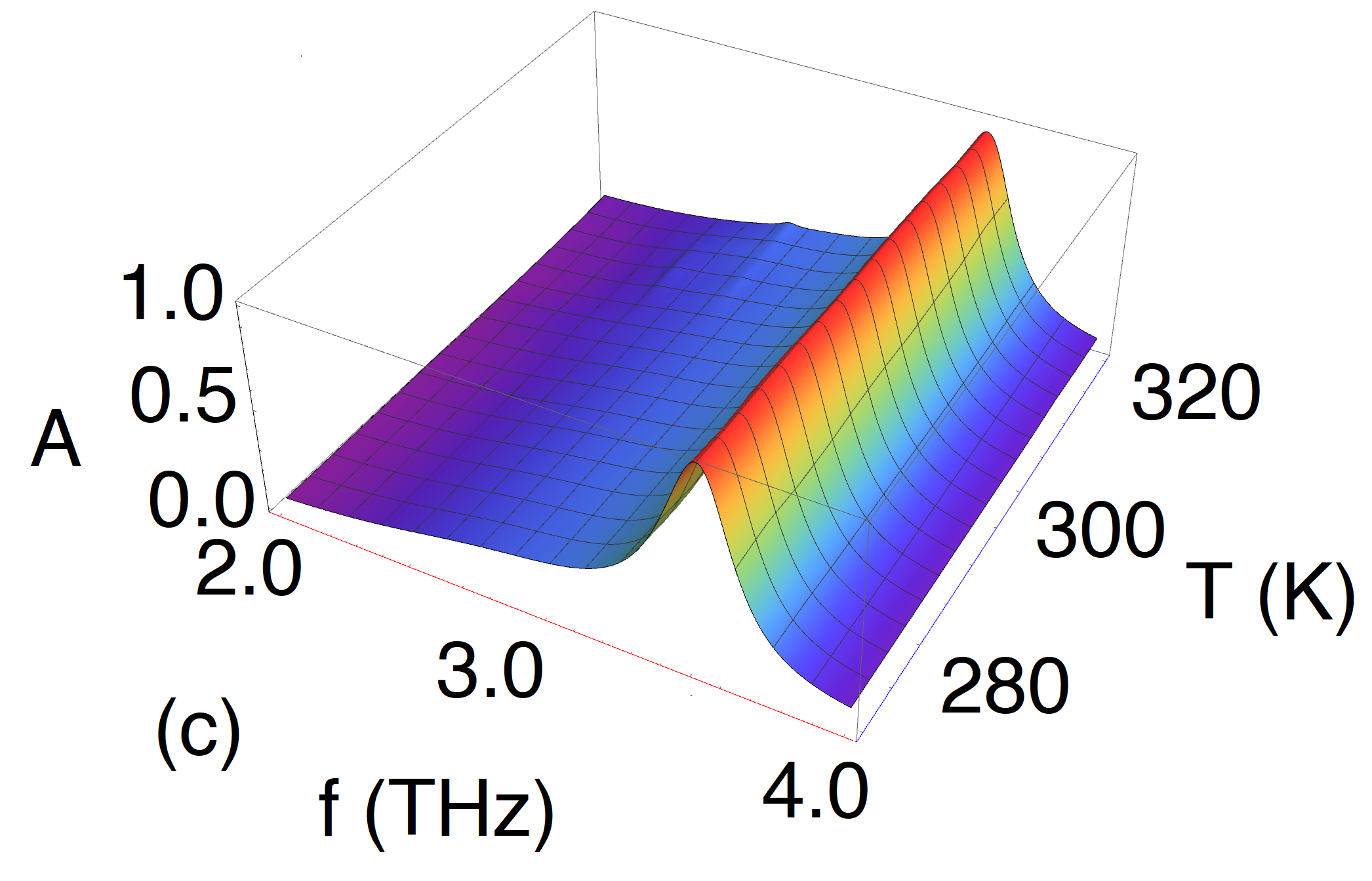}\\
\includegraphics[width=0.3 \columnwidth]{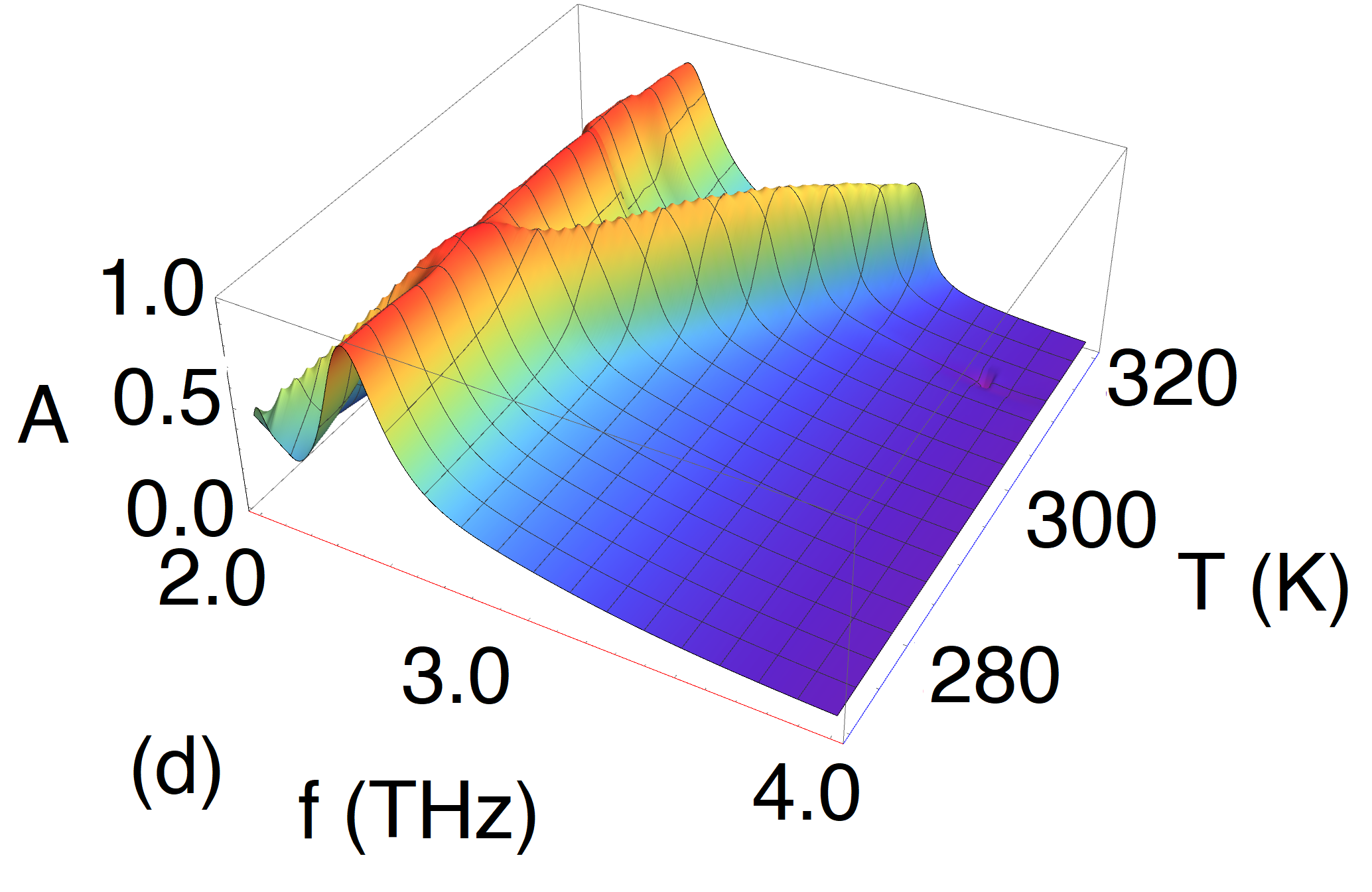}
\includegraphics[width=0.3 \columnwidth]{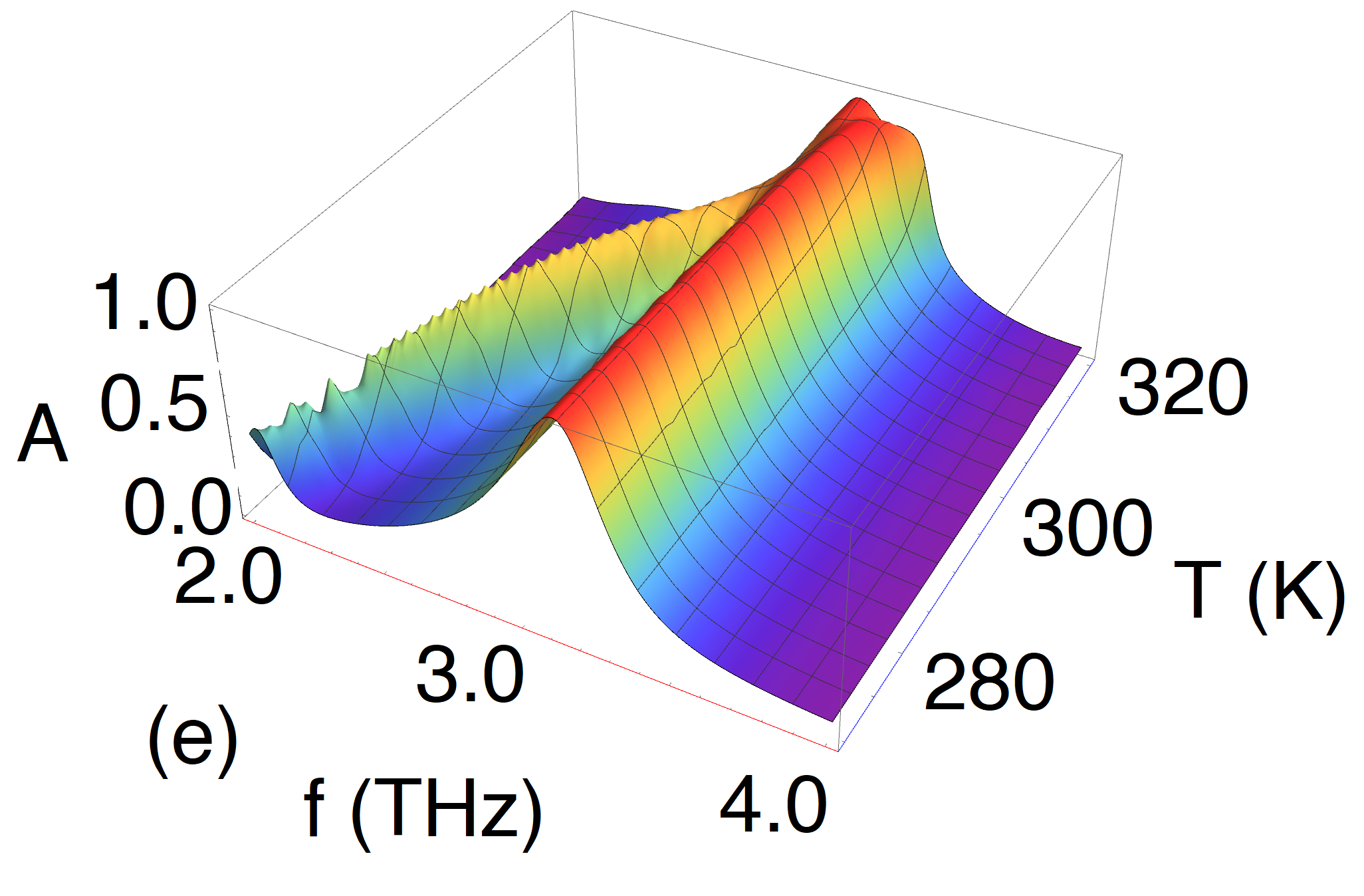}
\includegraphics[width=0.3 \columnwidth]{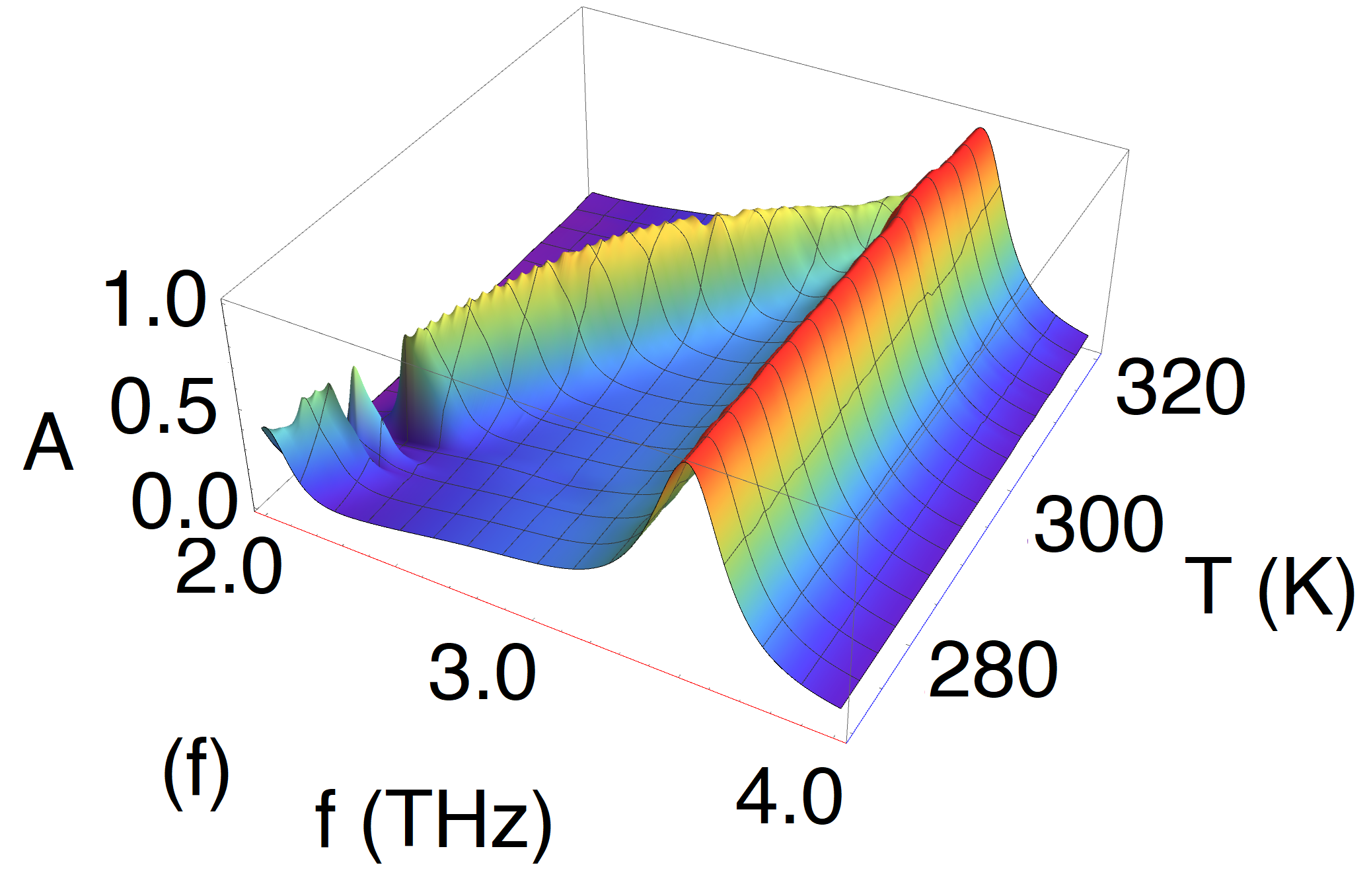}\\
\includegraphics[width=0.3 \columnwidth]{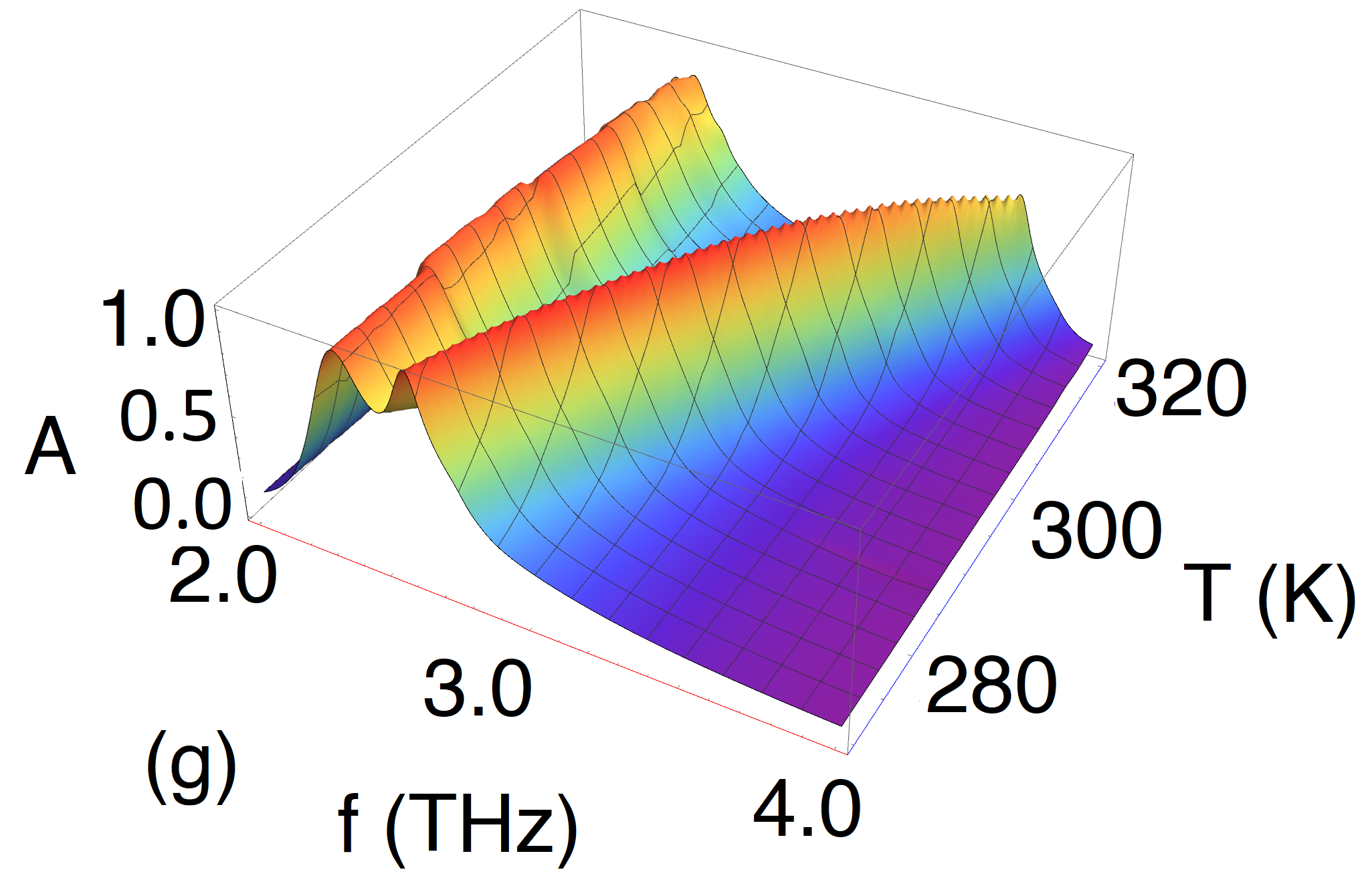}
\includegraphics[width=0.3 \columnwidth]{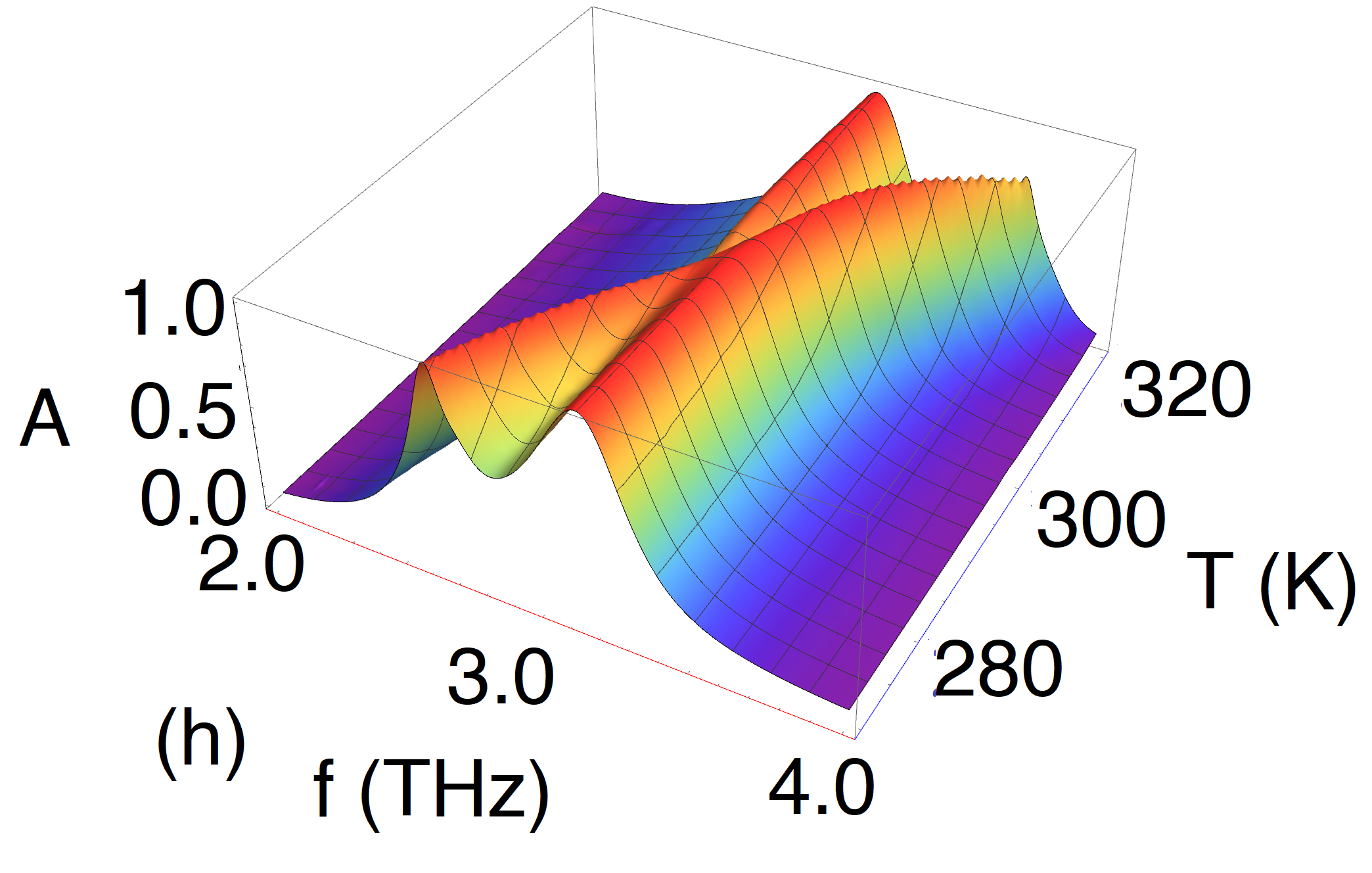}
\includegraphics[width=0.3 \columnwidth]{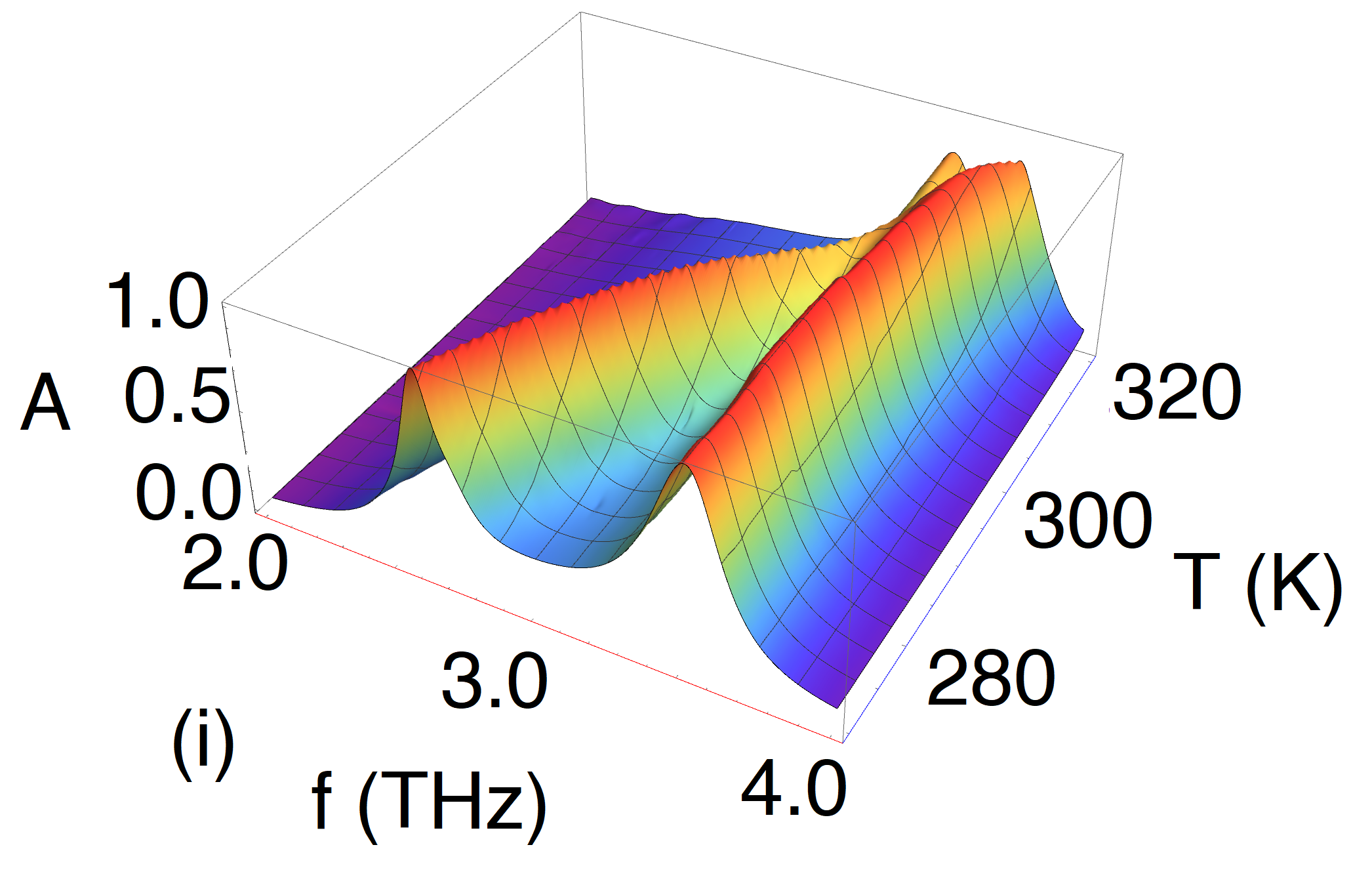}
}
\caption{Absorptance $A$ of the metasurface depicted via Fig.~\ref{geometry}b
as a function
of  $f\in[2,4]$~THz and $T\in[270,320]$~K, when  {$\varphi=0^\circ$}.
(a--c) $\Bdc=0$ with (a)
$\vert\Edc\vert=0.02$~\Vnm,
(b) 
$\vert\Edc\vert=0.05$~\Vnm, and
(c)
$\vert\Edc\vert=0.08$~\Vnm;
(d--f) $\vert\Bdc\vert=0.5$~T with 
(d) $\vert\Edc\vert=0.02$~\Vnm,
(e)  $\vert\Edc\vert=0.05$~\Vnm, and
(f) 
$\vert\Edc\vert=0.08$~\Vnm; and
(g--i)  $\vert\Bdc\vert=1$~T with (g)
$\vert\Edc\vert=0.02$~\Vnm,
(h)  
$\vert\Edc\vert=0.05$~\Vnm, and
(i)
$\vert\Edc\vert=0.08$~\Vnm. 
 \label{Abs2}
}

\end{figure}
\end{widetext}

 The introduction of $\vert\Bdc\vert>0$ of a sufficiently large
 magnitude creates a second peak-shaped feature in the absorptance
 spectrum, as is evident
 in Figs.~\ref{Abs2}d--i. Whereas the first feature remains almost temperature independent,
the second feature is strongly dependent on temperature.\\

\noindent \underline{1st peak-shaped feature}\\

\begin{figure}[ht]
\centering{
\includegraphics[width=0.3 \columnwidth]{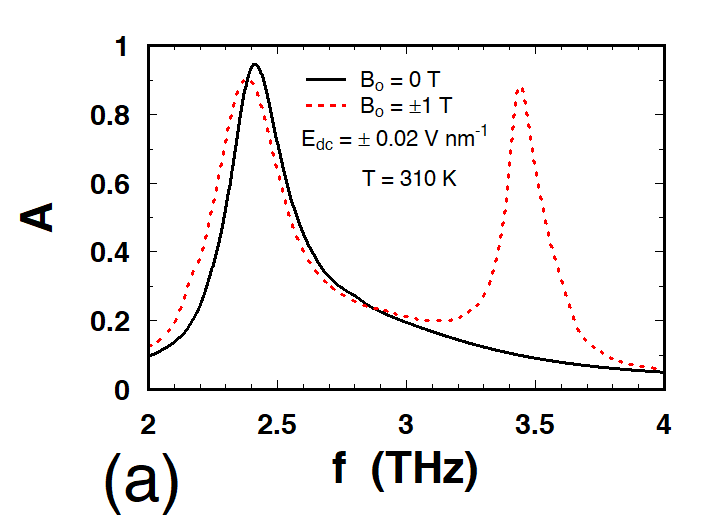}
\includegraphics[width=0.3 \columnwidth]{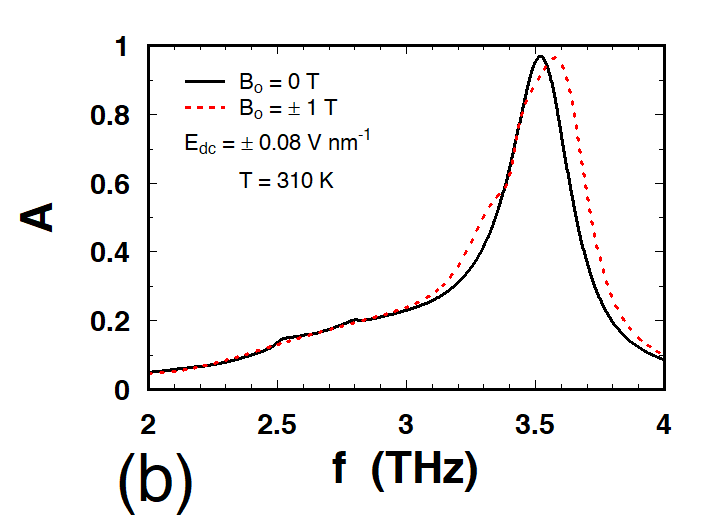}
}
\caption{Absorptance  $A$ of the metasurface depicted via Fig.~\ref{geometry}b
as a function
of  $f\in[2,4]$~THz for $\vert\Bdc\vert\in\left\{0,1\right\}$~T at $T=310$~K, when
 {$\varphi=0^\circ$} and
(a) $\vert\Edc\vert= 0.02$~\Vnm
and 
(b) $\vert\Edc\vert= 0.08$~\Vnm.
  \label{Abs3}
}

\end{figure}

Let us first discuss the peak-shaped feature that is very weakly  dependent on temperature.
A comparison of the three graphs in any column of Fig.~\ref{Abs2} shows that feature
is also weakly dependent on the quasistatic magnetic field. For $\vert\Edc\vert= 0.02$~\Vnm
and $T=310$~K, a comparison of Figs.~\ref{Abs2}a, \ref{Abs2}d, and \ref{Abs2}g shows that
the maximum-absorptance frequency  redshifts from $2.41$~THz
to $2.38$~THz at the rate of about
$0.03$~\THzT, as $\vert\Bdc\vert$ is increased from $0$ to $1$~T.
For $\vert\Edc\vert= 0.08$~\Vnm
and $T=310$~K, a comparison of Figs.~\ref{Abs2}c, \ref{Abs2}f, and \ref{Abs2}i shows that
the maximum-absorptance frequency  blueshifts from $3.52$~THz to $3.57$~THz
at the rate of about
$0.05$~\THzT, as $\vert\Bdc\vert$ is increased from $0$ to $1$~T. These spectral shifts
are evident in the absorption spectrums provided in Fig.~\ref{Abs3}. 

However,
if $\vert\Bdc\vert$ is fixed between $0$ and $1$~T, a quasistatic electric field
can be used to blueshift the maximum-absorptance frequency at the rate
of about $19.1$~\THzVnm.
Thus,  the 1st peak-shaped feature is controlled primarily
by  a quasistatic electric field that acts through
 the graphene-patched pixel.\\
 
 \noindent \underline{2nd peak-shaped feature}\\
 
 The second peak-shaped feature is absent for $\Bdc=0$ and it is also strongly dependent
 on temperature. Thus, it surely arises from the bicontrollability of the six InSb-patched
 pixels. The magnetic control modality becomes obvious on comparing Figs.~\ref{Abs2}d
 and \ref{Abs2}g, for both of which $\vert\Edc\vert=0.02$~\Vnm.
 In addition,  electric control  can be strongly exerted through the sole graphene-patched pixel,
 as is clear on comparing Figs.~\ref{Abs2}h
 and \ref{Abs2}i, for both of which $\vert\Bdc\vert=1$~T.

 \subsubsection{Example No. 1\label{ex1}} 
\begin{figure}[ht]
\centering{
\includegraphics[width=0.3 \columnwidth]{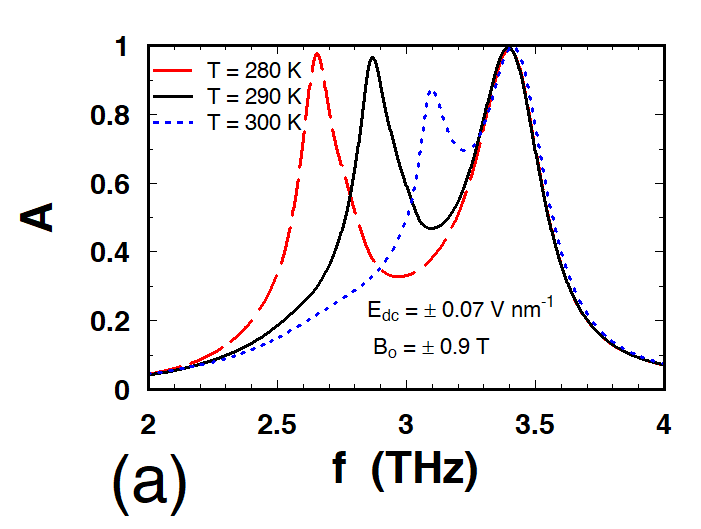}
\includegraphics[width=0.3 \columnwidth]{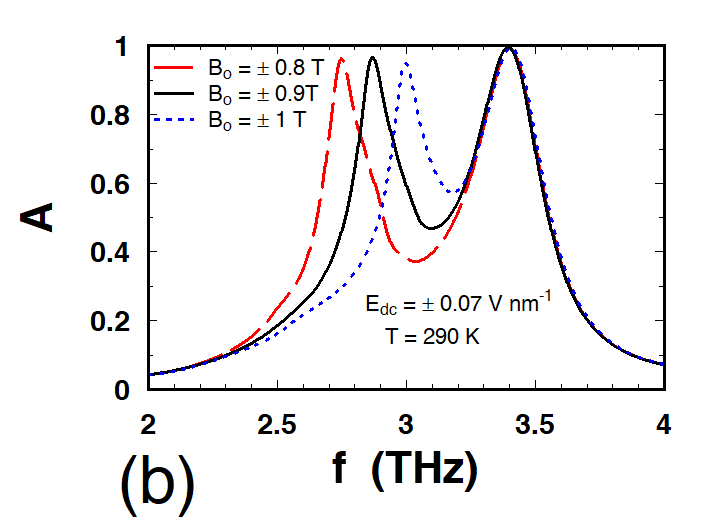}
\includegraphics[width=0.3 \columnwidth]{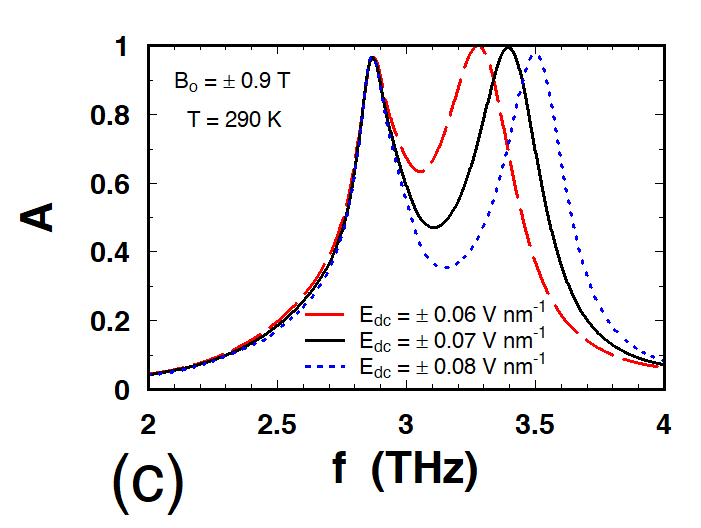} 
}
\caption{Absorptance  $A$ of the metasurface depicted via Fig.~\ref{geometry}b
as a function
of  $f\in[2,4]$~THz, when
(a) $T\in\left\{280, 290,300\right\}$~K, $\vert\Bdc\vert=0.9$~T, and $\vert\Edc\vert=0.07$~\Vnm;
(b) $T= 290$~K, $\vert\Bdc\vert\in\left\{0.8,0.9,1\right\}$~T, and $\vert\Edc\vert=0.07$~\Vnm;
and
(c) $T= 290$~K, $\vert\Bdc\vert=0.9$~T, and $\vert\Edc\vert\in\left\{0.06,0.07,0.08\right\}$~\Vnm.
 These data were computed for  {$\varphi=0^\circ$}.   
 \label{Abs4}
}

\end{figure}

 An example presented through
Fig.~\ref{Abs4} allows  appreciation of tricontrollability quite easily. Suppose
that the quiescent conditions for the operation of the
metasurface depicted via Fig.~\ref{geometry}b are as follows:
 $T= 290$~K, $\vert\Bdc\vert=0.9$~T, and $\vert\Edc\vert=0.07$~\Vnm. Then,
as shown through the black solid curves in Fig.~\ref{Abs4}, the absorptance spectrum
contains both peak-shaped features. The low-frequency
feature has a maximum-absorptance
frequency of    $2.87$~THz and
 FWHM bandwidth of  $0.28$~THz. The high-frequency
feature has a maximum-absorptance
frequency of   $3.39$~THz and
 FWHM bandwidth of $0.38$~THz.

 \noindent \underline{Thermal control modality}\\

If  $T$ is reduced to $280$~K from the quiescent point, the maximum-absorptance
frequency of the low-frequency feature redshifts to  $2.65$~THz and
the FWHM bandwidth alters to  $0.25$~THz  in Fig.~\ref{Abs4}a. In the same figure,
 the maximum-absorptance
frequency of the low-frequency feature changes to  $3.09$~THz and
the FWHM bandwidth to   $0.56$~THz, if  $T$
is increased to $300$~K from the quiescent point. Thus,
the maximum-absorptance frequency of the low-frequency feature blueshifts at the average rate
of     $0.022$~\THzK as the temperature increases by $20$~K.
The maximum absorptance diminishes
 from  $0.97$  at $T=280$~K to  $0.87$~THz at $T=300$~K.

For the high-frequency feature, the maximum-absorptance frequency is  $3.38$~THz
at $T=280$~K and  $3.41$~THz
at $T=300$~K. The FWHM bandwidth remains about 
 $0.35$~THz and the maximum absorptance does 
not decrease below  $0.985$, despite the temperature change. 
Thus,
the maximum-absorptance frequency of the high-frequency feature in Fig.~\ref{Abs4}a 
blueshifts at the average rate
of    $0.0015$~\THzK as the temperature increases by $20$~K.

We conclude that temperature can be used for: (i) coarse control of the low-frequency peak-shaped feature
and (ii) ultrafine control of the high-frequency peak-shaped feature. This control modality comes
from the six InSb-patched pixels in the meta-atoms of the chosen metasurface.\\

 \noindent \underline{Magnetic control modality}\\
 
 If  $\vert\Bdc\vert$ is reduced to $0.8$~T from the quiescent point, the maximum-absorptance
frequency of the low-frequency feature redshifts to $2.74$~THz and
the FWHM bandwidth changes to  $0.26$~THz  in Fig.~\ref{Abs4}b. On the other hand,
if  $\vert\Bdc\vert$ is increased to $1$~T from the quiescent point, the maximum-absorptance
frequency of the low-frequency feature blueshifts to  $2.99$~THz and
the FWHM bandwidth changes to   $0.67$~THz. Thus,
the maximum-absorptance frequency of the low-frequency feature  blueshifts at the average rate
of    $1.2$~\THzT as the quasistatic magnetic field's magnitude increases by $0.2$~T,
the maximum absorptance remaining about  $0.94$ despite that  increase.

For the high-frequency feature, the maximum-absorptance frequency is $3.38$~THz
at $\vert\Bdc\vert=0.8$~T and $3.40$~THz
at $\vert\Bdc\vert=1$~T in  Fig.~\ref{Abs4}b. The FWHM bandwidth is
$0.36$~THz at $\vert\Bdc\vert=0.8$~T and 
$0.67$~THz at $\vert\Bdc\vert=1$~T. The maximum
absorptance remains  in excess of $0.99$, despite the increase in the
magnitude of the quasistatic magnetic field's magnitude   by $0.2$~T.
Thus,
the maximum-absorptance frequency of the high-frequency feature  blueshifts at the average rate
of   $0.01$~\THzT as  $\vert\Bdc\vert$ increases by $0.2$~T  in Fig.~\ref{Abs4}b.

Accordingly,   a quasistatic magnetic field oriented tangentially to the chosen metasurface can be used for: (i)
coarse control of the low-frequency peak-shaped feature and (ii)
ultrafine control of the high-frequency peak-shaped feature. This control modality comes
from the six InSb-patched pixels in the meta-atoms. 

{Thus, if the magnet providing magnetic control  becomes unavailable, the thermal control modality can be used. Conversely, if the heater/cooler unit providing thermal control becomes unavailable, the magnetic control modality can be used. If both modalities are available, both can be operationally hybridized to reduce the energy expended to run the control modalities.}
\\

\noindent \underline{Electrical control modality}\\

If  $\vert\Edc\vert$ is reduced to $0.06$~\Vnm from the quiescent point, the maximum-absorptance
frequency of the low-frequency feature  blueshifts slightly to $2.872$~THz and
the FWHM bandwidth changes to  $0.69$~THz  in Fig.~\ref{Abs4}c. In the same figure,
 the maximum-absorptance
frequency of the low-frequency feature  redshifts to  $2.86$~THz and
the FWHM bandwidth alters to  $0.24$~THz, if  $\vert\Edc\vert$ is increased to $0.08$~\Vnm from the quiescent point. Thus,
the maximum-absorptance frequency of the low-frequency feature  redshifts at the average rate
of    $0.5$~{\THzVnm} as the magnitude of the quasistatic electric field increases by $0.02$~\Vnm.
The maximum absorptance remains in excess of  $0.96$. 

For the high-frequency feature, the maximum-absorptance frequency is  $3.27$~THz
at $\vert\Edc\vert=0.06$~\Vnm and  $3.49$~THz
at $\vert\Edc\vert=0.08$~\Vnm. The FWHM bandwidth is
$0.69$~THz at $\vert\Edc\vert=0.06$~\Vnm and 
$0.32$~THz at $\vert\Edc\vert=0.08$~\Vnm. The maximum
absorptance remains in excess of 
$0.98$,  even though the magnitude of the quasistatic electric field is increased by $0.02$~\Vnm.
Thus,
the maximum-absorptance frequency of the high-frequency feature in Fig.~\ref{Abs4}c 
blueshifts at the average rate
of   $11$~{\THzVnm} as $\vert\Edc\vert$ is increased by $0.02$~\Vnm.

Accordingly, a quasistatic electric field oriented normally to the chosen metasurface can be used for: (i) ultrafine control of the low-frequency peak-shaped feature
and (ii) coarse control of the high-frequency peak-shaped feature. This control modality comes
from the sole graphene-patched pixel in the meta-atoms.\\

 \subsubsection{Example No. 2 \label{ex2}} 
\begin{figure}[ht]
\centering{
\includegraphics[width=0.3 \columnwidth]{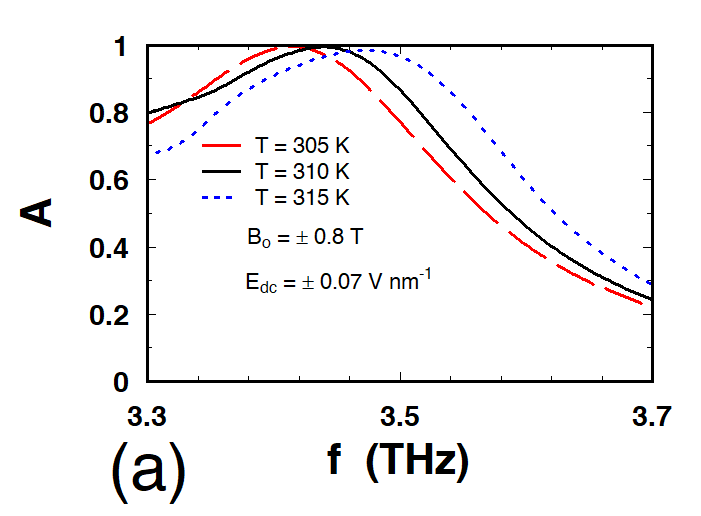}
\includegraphics[width=0.3 \columnwidth]{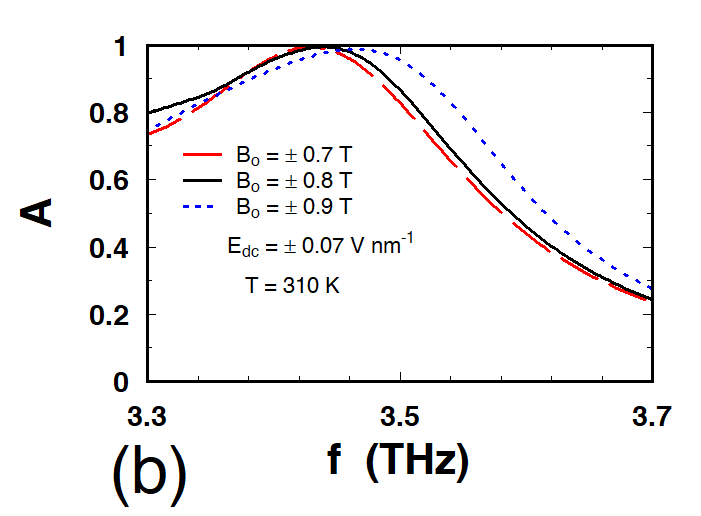}
\includegraphics[width=0.3 \columnwidth]{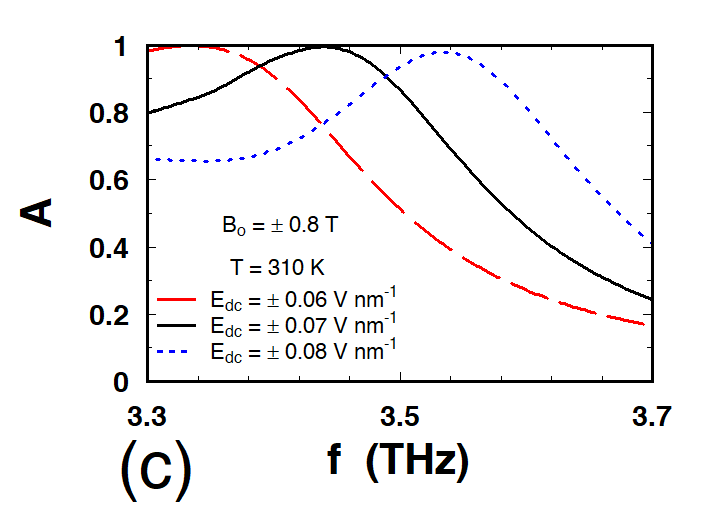} 
}
\caption{Absorptance  $A$ of the metasurface depicted via Fig.~\ref{geometry}b
as a function
of  $f\in[3.3,3.7]$~THz, when
(a) $T\in\left\{305,310,315\right\}$~K, $\vert\Bdc\vert=0.8$~T, and $\vert\Edc\vert=0.07$~\Vnm;
(b) $T= 310$~K, $\vert\Bdc\vert\in\left\{0.7,0.8,0.9\right\}$~T, and $\vert\Edc\vert=0.07$~\Vnm;
and
(c) $T= 310$~K, $\vert\Bdc\vert=0.8$~T, and $\vert\Edc\vert\in\left\{0.06,0.07,0.08\right\}$~\Vnm.
 These data were computed for  {$\varphi=0^\circ$}.   
 \label{Abs5}
}

\end{figure}

 In Example No. 1 (Sec.~3.\ref{ex1}), either one or two  modalities provide coarse control,
 whereas the remaining   modalities provide fine control, of a
 maximum-absorptance frequency. Through the absorptance
 spectrums in Figs.~\ref{Abs5}a--c, we present  an example
 wherein all three modalities provide comparable control.
 
The quiescent conditions for the operation of the
metasurface depicted via Fig.~\ref{geometry}b are chosen for Example No.~2
as follows:
 $T= 310$~K, $\vert\Bdc\vert=0.8$~T, and $\vert\Edc\vert=0.07$~\Vnm.
 According to the black
 solid curves in Fig.~\ref{Abs5}, the maximum absorptance is $0.99$,
 the maximum-absorptance frequency is $3.44$~THz, and the FWHM
 bandwidth is $0.47$~THz.

When  $T$ is changed from $305$~K to $315$~K but both  $\vert\Bdc\vert=0.8$~T and $\vert\Edc\vert=0.07$~\Vnm~remain invariant,
 Fig.~\ref{Abs5}a shows that the maximum-absorptance frequency blueshifts
 from $3.41$~THz to $3.47$~THz at the approximate rate of $0.006$~\THzK, and
 the FWHM bandwidth reduces from $0.54$~THz to $0.45$~THz, but the
 maximum absorptance remains at least $0.98$.
 
  When  $\vert \Bdc\vert$ is changed from $0.7$~T to $0.9$~T but 
  both $T= 310$~K  and $\vert\Edc\vert=0.07$~\Vnm~are held fixed,
 Fig.~\ref{Abs5}b shows that the maximum-absorptance frequency blueshifts
 from $3.43$~THz to $3.46$~THz at the approximate rate of $0.15$~\THzT, and
 the FWHM bandwidth reduces from $0.51$~THz to $0.44$~THz, but the
 maximum absorptance remains at least $0.98$.
 
 Finally, when  $\vert \Edc\vert$ is changed from $0.06$~\Vnm~  to $0.08$~\Vnm~but 
  both $T= 310$~K  and $\vert\Bdc\vert=0.8$~T are held fixed,
 Fig.~\ref{Abs5}c shows that the maximum-absorptance frequency blueshifts
 from $3.33$~THz to $3.53$~THz at the approximate rate of $10$~{\THzVnm}, and
 the FWHM bandwidth reduces from  $0.51$~THz to $0.49$~THz, but the
 maximum absorptance remains at least $0.98$.
 
 Thus, either of the three control modalities can be employed to tune the maximum-absorptance
 frequency with comparable degree of control. 
 
 \subsubsection{Mechanical control modality}
 There is a fourth control modality available too!
 Suppose that the polarization angle $\varphi$ is changed from $0^\circ$. 
 The absorptance spectrum
 will change.
 
\begin{figure}[ht]
\centering{
\includegraphics[width=0.3 \columnwidth]{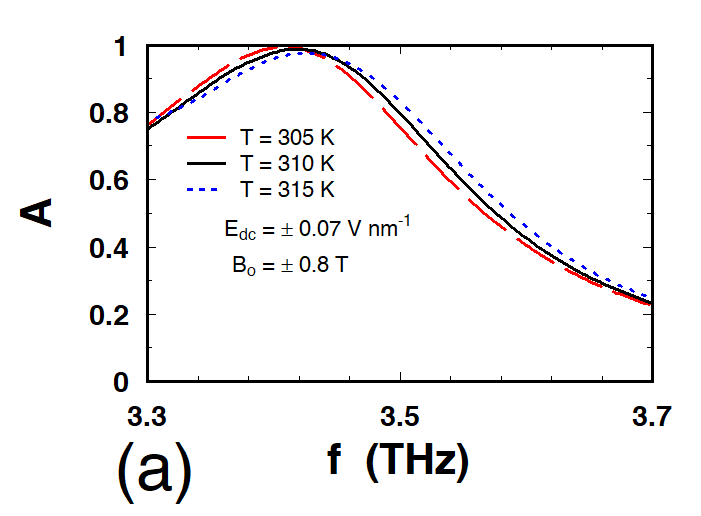}
\includegraphics[width=0.3 \columnwidth]{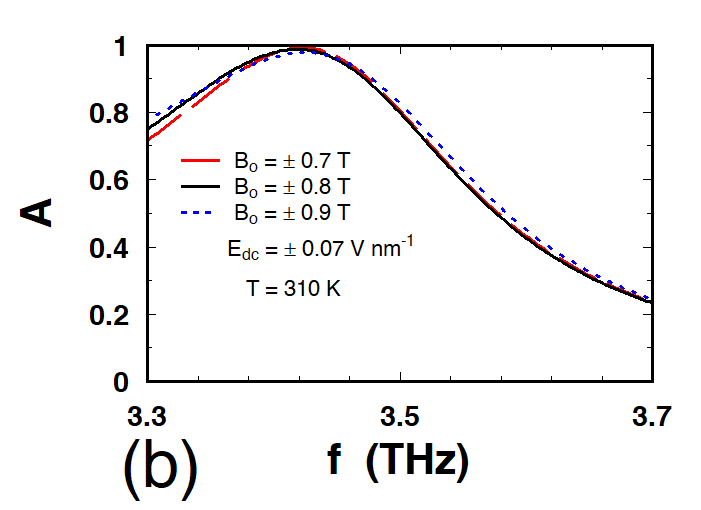}
\includegraphics[width=0.3 \columnwidth]{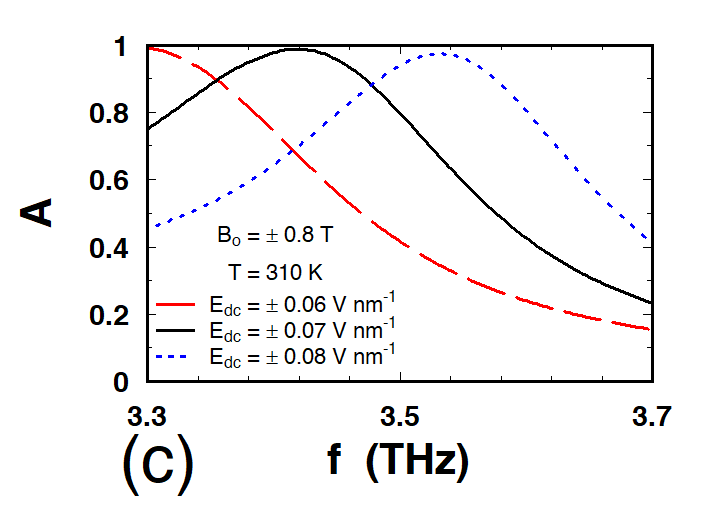} 
}
\caption{Same as Fig.~\ref{Abs5}, except for    {$\varphi=90^\circ$}.   
 \label{Abs6}
}

\end{figure}

 As an example, the absorptance spectrums  drawn for Example No. 2
  in Sec.~\ref{ex2} were recalculated for $\varphi=90^\circ$.  With
  the quiescent conditions
  $T= 310$~K, $\vert\Bdc\vert=0.8$~T, and $\vert\Edc\vert=0.07$~\Vnm~
  being the same as with $\varphi=0^\circ$ (Fig.~\ref{Abs5}),  the maximum absorptance remains $0.99$ but
 the maximum-absorptance frequency changes to $3.42$~THz and the FWHM
 bandwidth to $0.43$~THz  when $\varphi$ is changed to $90^\circ$ (Fig.~\ref{Abs6}). This amounts
 to fine control exerted through rotation of the incident plane wave by $90^\circ$.
 Similar changes occur when $T$, $\vert\Bdc\vert$, and $\vert\Edc\vert$ are
 changed additionally. Therefore, {the lack of rotational invariance of   the metasurface depicted via Fig.~\ref{geometry}b about the $z$ axis} 
 facilitates a mechanical control modality.
 
\section{Concluding Remarks} \label{conc}
The piexelated-metasurface approach \cite{Lakhtakia} was inspired by examples
from biology in order to design metasurfaces whose electromagnetic response
characteristics can be controlled by several different agencies. Theory had previously shown
\cite{Chiadini}
that thermal
and magnetic control modalities would be effective for a bicontrollable metasurface
comprising pixels patched with CdTe for thermal control and InAs for magnetic control.
In this paper, we have presented a tricontrollable metasurface comprising
pixels patched with graphene for electrical control
and InSb for thermal and magnetic control. The {lack of rotational invariance} of the optimal
meta-atom comprising unpatched, graphene-patched, and InSb-patched pixels
adds mechanical rotation as the fourth control modality. With  {microlithographic techniques
entailing physical and chemical vapor depositions over a series of masks combined
with etching \cite{FranssilaBook,LakhtakiaBook}
available} to fabricate such metasurfaces, we expect that our theoretical efforts will
inspire experimentalist colleagues.

\vspace{0.5cm}
\noindent {\bf Acknowledgments.}  {AL and PKJ thank the Visiting Advanced Joint Research (VAJRA) program of the  Department of Science and Technology (Government of India) for initially funding their collaborative research. AL thanks the Charles Godfrey Binder Endowment at Penn State for ongoing support of his research activities and the Otto M{\o}nsted Foundation for generous financial support during an extended stay at the Danish Technical University.}

\end{document}